\documentclass[journal,10pt,twoside,twocolumn]{IEEEtran}

\usepackage{mleftright}
\usepackage{cite}
\usepackage[T1]{fontenc}
\usepackage{graphicx}
\usepackage{mathtools}
\usepackage{amssymb}
\usepackage{amsmath}
\usepackage{paralist}
\usepackage{subfigure}
\usepackage{booktabs} 
\usepackage{algorithm}
\usepackage{algorithmic}
\usepackage{amsthm}
\usepackage{multirow}
\usepackage{microtype}
\usepackage{balance}
\usepackage{amssymb}
\usepackage{amsfonts}
\usepackage{mathrsfs}
\usepackage{xspace}
\usepackage{bm}
\usepackage{upgreek}

\newcommand{\safemath}[2]{\newcommand{#1}{\ensuremath{#2}\xspace}}

\safemath{\bma}{\mathbf{a}}
\safemath{\bmb}{\mathbf{b}}
\safemath{\bmc}{\mathbf{c}}
\safemath{\bmd}{\mathbf{d}}
\safemath{\bme}{\mathbf{e}}
\safemath{\bmf}{\mathbf{f}}
\safemath{\bmg}{\mathbf{g}}
\safemath{\bmh}{\mathbf{h}}
\safemath{\bmi}{\mathbf{i}}
\safemath{\bmj}{\mathbf{j}}
\safemath{\bmk}{\mathbf{k}}
\safemath{\bml}{\mathbf{l}}
\safemath{\bmm}{\mathbf{m}}
\safemath{\bmn}{\mathbf{n}}
\safemath{\bmo}{\mathbf{o}}
\safemath{\bmp}{\mathbf{p}}
\safemath{\bmq}{\mathbf{q}}
\safemath{\bmr}{\mathbf{r}}
\safemath{\bms}{\mathbf{s}}
\safemath{\bmt}{\mathbf{t}}
\safemath{\bmu}{\mathbf{u}}
\safemath{\bmv}{\mathbf{v}}
\safemath{\bmw}{\mathbf{w}}
\safemath{\bmx}{\mathbf{x}}
\safemath{\bmy}{\mathbf{y}}
\safemath{\bmz}{\mathbf{z}}
\safemath{\bmzero}{\mathbf{0}}
\safemath{\bmone}{\mathbf{1}}

\bmdefine{\biad}{a}
\bmdefine{\bibd}{b}
\bmdefine{\bicd}{c}
\bmdefine{\bidd}{d}
\bmdefine{\bied}{e}
\bmdefine{\bifd}{f}
\bmdefine{\bigd}{g}
\bmdefine{\bihd}{h}
\bmdefine{\biid}{i}
\bmdefine{\bijd}{j}
\bmdefine{\bikd}{k}
\bmdefine{\bild}{l}
\bmdefine{\bimd}{m}
\bmdefine{\bind}{n}
\bmdefine{\biod}{o}
\bmdefine{\bipd}{p}
\bmdefine{\biqd}{q}
\bmdefine{\bird}{r}
\bmdefine{\bisd}{s}
\bmdefine{\bitd}{t}
\bmdefine{\biud}{u}
\bmdefine{\bivd}{v}
\bmdefine{\biwd}{w}
\bmdefine{\bixd}{x}
\bmdefine{\biyd}{y}
\bmdefine{\bizd}{z}

\bmdefine{\bixid}{\xi}
\bmdefine{\bilambdad}{\lambda}
\bmdefine{\bimud}{\mu}
\bmdefine{\bithetad}{\theta}
\bmdefine{\biphid}{\phi}
\bmdefine{\bideltad}{\delta}

\safemath{\bmia}{\biad}
\safemath{\bmib}{\bibd}
\safemath{\bmic}{\bicd}
\safemath{\bmid}{\bidd}
\safemath{\bmie}{\bied}
\safemath{\bmif}{\bifd}
\safemath{\bmig}{\bigd}
\safemath{\bmih}{\bihd}
\safemath{\bmii}{\biid}
\safemath{\bmij}{\bijd}
\safemath{\bmik}{\bikd}
\safemath{\bmil}{\bild}
\safemath{\bmim}{\bimd}
\safemath{\bmin}{\bind}
\safemath{\bmio}{\biod}
\safemath{\bmip}{\bipd}
\safemath{\bmiq}{\biqd}
\safemath{\bmir}{\bird}
\safemath{\bmis}{\bisd}
\safemath{\bmit}{\bitd}
\safemath{\bmiu}{\biud}
\safemath{\bmiv}{\bivd}
\safemath{\bmiw}{\biwd}
\safemath{\bmix}{\bixd}
\safemath{\bmiy}{\biyd}
\safemath{\bmiz}{\bizd}

\safemath{\bmxi}{\bixid}
\safemath{\bmlambda}{\bilambdad}
\safemath{\bmmu}{\bimud}
\safemath{\bmtheta}{\bithetad}
\safemath{\bmphi}{\biphid}
\safemath{\bmdelta}{\bideltad}

\safemath{\bA}{\mathbf{A}}
\safemath{\bB}{\mathbf{B}}
\safemath{\bC}{\mathbf{C}}
\safemath{\bD}{\mathbf{D}}
\safemath{\bE}{\mathbf{E}}
\safemath{\bF}{\mathbf{F}}
\safemath{\bG}{\mathbf{G}}
\safemath{\bH}{\mathbf{H}}
\safemath{\bI}{\mathbf{I}}
\safemath{\bJ}{\mathbf{J}}
\safemath{\bK}{\mathbf{K}}
\safemath{\bL}{\mathbf{L}}
\safemath{\bM}{\mathbf{M}}
\safemath{\bN}{\mathbf{N}}
\safemath{\bO}{\mathbf{O}}
\safemath{\bP}{\mathbf{P}}
\safemath{\bQ}{\mathbf{Q}}
\safemath{\bR}{\mathbf{R}}
\safemath{\bS}{\mathbf{S}}
\safemath{\bT}{\mathbf{T}}
\safemath{\bU}{\mathbf{U}}
\safemath{\bV}{\mathbf{V}}
\safemath{\bW}{\mathbf{W}}
\safemath{\bX}{\mathbf{X}}
\safemath{\bY}{\mathbf{Y}}
\safemath{\bZ}{\mathbf{Z}}

\safemath{\bZero}{\mathbf{0}}
\safemath{\bOne}{\mathbf{1}}
\safemath{\bDelta}{\mathbf{\Delta}}
\safemath{\bLambda}{\mathbf{\UpLambda}}
\safemath{\bPhi}{\mathbf{\Upphi}}
\safemath{\bSigma}{\mathbf{\Upsigma}}
\safemath{\bOmega}{\mathbf{\Upomega}}
\safemath{\bTheta}{\mathbf{\Uptheta}}

\bmdefine{\biAd}{A}
\bmdefine{\biBd}{B}
\bmdefine{\biCd}{C}
\bmdefine{\biDd}{D}
\bmdefine{\biEd}{E}
\bmdefine{\biFd}{F}
\bmdefine{\biGd}{G}
\bmdefine{\biHd}{H}
\bmdefine{\biId}{I}
\bmdefine{\biJd}{J}
\bmdefine{\biKd}{K}
\bmdefine{\biLd}{L}
\bmdefine{\biMd}{M}
\bmdefine{\biOd}{N}
\bmdefine{\biPd}{O}
\bmdefine{\biQd}{P}
\bmdefine{\biRd}{R}
\bmdefine{\biSd}{S}
\bmdefine{\biTd}{T}
\bmdefine{\biUd}{U}
\bmdefine{\biVd}{V}
\bmdefine{\biWd}{W}
\bmdefine{\biXd}{X}
\bmdefine{\biYd}{Y}
\bmdefine{\biZd}{Z}

\bmdefine{\biDelta}{\Delta}
\bmdefine{\biLambda}{\Lambda}
\bmdefine{\biPhi}{\Phi}
\bmdefine{\biSigma}{\Sigma}
\bmdefine{\biOmega}{\Omega}
\bmdefine{\biTheta}{\Theta}

\safemath{\bimA}{\biAd}
\safemath{\bimB}{\biBd}
\safemath{\bimC}{\biCd}
\safemath{\bimD}{\biDd}
\safemath{\bimE}{\biEd}
\safemath{\bimF}{\biFd}
\safemath{\bimG}{\biGd}
\safemath{\bimH}{\biHd}
\safemath{\bimI}{\biId}
\safemath{\bimJ}{\biJd}
\safemath{\bimK}{\biKd}
\safemath{\bimL}{\biLd}
\safemath{\bimM}{\biMd}
\safemath{\bimN}{\biNd}
\safemath{\bimO}{\biOd}
\safemath{\bimP}{\biPd}
\safemath{\bimQ}{\biQd}
\safemath{\bimR}{\biRd}
\safemath{\bimS}{\biSd}
\safemath{\bimT}{\biTd}
\safemath{\bimU}{\biUd}
\safemath{\bimV}{\biVd}
\safemath{\bimW}{\biWd}
\safemath{\bimX}{\biXd}
\safemath{\bimY}{\biYd}
\safemath{\bimZ}{\biZd}

\safemath{\bimDelta}{\biDelta}
\safemath{\bimLambda}{\biLambda}
\safemath{\bimPhi}{\biPhi}
\safemath{\bimSigma}{\biSigma}
\safemath{\bimOmega}{\biOmega}
\safemath{\bimTheta}{\biTheta}

\safemath{\setA}{\mathcal{A}}
\safemath{\setB}{\mathcal{B}}
\safemath{\setC}{\mathcal{C}}
\safemath{\setD}{\mathcal{D}}
\safemath{\setE}{\mathcal{E}}
\safemath{\setF}{\mathcal{F}}
\safemath{\setG}{\mathcal{G}}
\safemath{\setH}{\mathcal{H}}
\safemath{\setI}{\mathcal{I}}
\safemath{\setJ}{\mathcal{J}}
\safemath{\setK}{\mathcal{K}}
\safemath{\setL}{\mathcal{L}}
\safemath{\setM}{\mathcal{M}}
\safemath{\setN}{\mathcal{N}}
\safemath{\setO}{\mathcal{O}}
\safemath{\setP}{\mathcal{P}}
\safemath{\setQ}{\mathcal{Q}}
\safemath{\setR}{\mathcal{R}}
\safemath{\setS}{\mathcal{S}}
\safemath{\setT}{\mathcal{T}}
\safemath{\setU}{\mathcal{U}}
\safemath{\setV}{\mathcal{V}}
\safemath{\setW}{\mathcal{W}}
\safemath{\setX}{\mathcal{X}}
\safemath{\setY}{\mathcal{Y}}
\safemath{\setZ}{\mathcal{Z}}
\safemath{\emptySet}{\varnothing}

\safemath{\colA}{\mathscr{A}}
\safemath{\colB}{\mathscr{B}}
\safemath{\colC}{\mathscr{C}}
\safemath{\colD}{\mathscr{D}}
\safemath{\colE}{\mathscr{E}}
\safemath{\colF}{\mathscr{F}}
\safemath{\colG}{\mathscr{G}}
\safemath{\colH}{\mathscr{H}}
\safemath{\colI}{\mathscr{I}}
\safemath{\colJ}{\mathscr{J}}
\safemath{\colK}{\mathscr{K}}
\safemath{\colL}{\mathscr{L}}
\safemath{\colM}{\mathscr{M}}
\safemath{\colN}{\mathscr{N}}
\safemath{\colO}{\mathscr{O}}
\safemath{\colP}{\mathscr{P}}
\safemath{\colQ}{\mathscr{Q}}
\safemath{\colR}{\mathscr{R}}
\safemath{\colS}{\mathscr{S}}
\safemath{\colT}{\mathscr{T}}
\safemath{\colU}{\mathscr{U}}
\safemath{\colV}{\mathscr{V}}
\safemath{\colW}{\mathscr{W}}
\safemath{\colX}{\mathscr{X}}
\safemath{\colY}{\mathscr{Y}}
\safemath{\colZ}{\mathscr{Z}}

\safemath{\opA}{\mathbb{A}}
\safemath{\opB}{\mathbb{B}}
\safemath{\opC}{\mathbb{C}}
\safemath{\opD}{\mathbb{D}}
\safemath{\opE}{\mathbb{E}}
\safemath{\opF}{\mathbb{F}}
\safemath{\opG}{\mathbb{G}}
\safemath{\opH}{\mathbb{H}}
\safemath{\opI}{\mathbb{I}}
\safemath{\opJ}{\mathbb{J}}
\safemath{\opK}{\mathbb{K}}
\safemath{\opL}{\mathbb{L}}
\safemath{\opM}{\mathbb{M}}
\safemath{\opN}{\mathbb{N}}
\safemath{\opO}{\mathbb{O}}
\safemath{\opP}{\mathbb{P}}
\safemath{\opQ}{\mathbb{Q}}
\safemath{\opR}{\mathbb{R}}
\safemath{\opS}{\mathbb{S}}
\safemath{\opT}{\mathbb{T}}
\safemath{\opU}{\mathbb{U}}
\safemath{\opV}{\mathbb{V}}
\safemath{\opW}{\mathbb{W}}
\safemath{\opX}{\mathbb{X}}
\safemath{\opY}{\mathbb{Y}}
\safemath{\opZ}{\mathbb{Z}}
\safemath{\opZero}{\mathbb{O}}
\safemath{\identityop}{\opI}

\safemath{\veca}{\bma}
\safemath{\vecb}{\bmb}
\safemath{\vecc}{\bmc}
\safemath{\vecd}{\bmd}
\safemath{\vece}{\bme}
\safemath{\vecf}{\bmf}
\safemath{\vecg}{\bmg}
\safemath{\vech}{\bmh}
\safemath{\veci}{\bmi}
\safemath{\vecj}{\bmj}
\safemath{\veck}{\bmk}
\safemath{\vecl}{\bml}
\safemath{\vecm}{\bmm}
\safemath{\vecn}{\bmn}
\safemath{\veco}{\bmo}
\safemath{\vecp}{\bmp}
\safemath{\vecq}{\bmq}
\safemath{\vecr}{\bmr}
\safemath{\vecs}{\bms}
\safemath{\vect}{\bmt}
\safemath{\vecu}{\bmu}
\safemath{\vecv}{\bmv}
\safemath{\vecw}{\bmw}
\safemath{\vecx}{\bmx}
\safemath{\vecy}{\bmy}
\safemath{\vecz}{\bmz}

\safemath{\veczero}{\bmzero}
\safemath{\vecone}{\bmone}
\safemath{\vecxi}{\bmxi}
\safemath{\veclambda}{\bmlambda}
\safemath{\vecmu}{\bmmu}
\safemath{\vectheta}{\bmtheta}
\safemath{\vecphi}{\bmphi}
\safemath{\vecdelta}{\bmdelta}

\safemath{\matA}{\bA}
\safemath{\matB}{\bB}
\safemath{\matC}{\bC}
\safemath{\matD}{\bD}
\safemath{\matE}{\bE}
\safemath{\matF}{\bF}
\safemath{\matG}{\bG}
\safemath{\matH}{\bH}
\safemath{\matI}{\bI}
\safemath{\matJ}{\bJ}
\safemath{\matK}{\bK}
\safemath{\matL}{\bL}
\safemath{\matM}{\bM}
\safemath{\matN}{\bN}
\safemath{\matO}{\bO}
\safemath{\matP}{\bP}
\safemath{\matQ}{\bQ}
\safemath{\matR}{\bR}
\safemath{\matS}{\bS}
\safemath{\matT}{\bT}
\safemath{\matU}{\bU}
\safemath{\matV}{\bV}
\safemath{\matW}{\bW}
\safemath{\matX}{\bX}
\safemath{\matY}{\bY}
\safemath{\matZ}{\bZ}
\safemath{\matzero}{\bmzero}

\safemath{\matDelta}{\bDelta}
\safemath{\matLambda}{\bLambda}
\safemath{\matPhi}{\bPhi}
\safemath{\matSigma}{\bSigma}
\safemath{\matOmega}{\bOmega}
\safemath{\matTheta}{\bTheta}

\safemath{\matidentity}{\matI}
\safemath{\matone}{\matO}

\safemath{\rnda}{A}
\safemath{\rndb}{B}
\safemath{\rndc}{C}
\safemath{\rndd}{D}
\safemath{\rnde}{E}
\safemath{\rndf}{F}
\safemath{\rndg}{G}
\safemath{\rndh}{H}
\safemath{\rndi}{I}
\safemath{\rndj}{J}
\safemath{\rndk}{K}
\safemath{\rndl}{L}
\safemath{\rndm}{M}
\safemath{\rndn}{N}
\safemath{\rndo}{O}
\safemath{\rndp}{P}
\safemath{\rndq}{Q}
\safemath{\rndr}{R}
\safemath{\rnds}{S}
\safemath{\rndt}{T}
\safemath{\rndu}{U}
\safemath{\rndv}{V}
\safemath{\rndw}{W}
\safemath{\rndx}{X}
\safemath{\rndy}{Y}
\safemath{\rndz}{Z}

\safemath{\rveca}{\bimA}
\safemath{\rvecb}{\bimB}
\safemath{\rvecc}{\bimC}
\safemath{\rvecd}{\bimD}
\safemath{\rvece}{\bimE}
\safemath{\rvecf}{\bimF}
\safemath{\rvecg}{\bimG}
\safemath{\rvech}{\bimH}
\safemath{\rveci}{\bimI}
\safemath{\rvecj}{\bimJ}
\safemath{\rveck}{\bimK}
\safemath{\rvecl}{\bimL}
\safemath{\rvecm}{\bimM}
\safemath{\rvecn}{\bimN}
\safemath{\rveco}{\bomO}
\safemath{\rvecp}{\bimP}
\safemath{\rvecq}{\bimQ}
\safemath{\rvecr}{\bimR}
\safemath{\rvecs}{\bimS}
\safemath{\rvect}{\bimT}
\safemath{\rvecu}{\bimU}
\safemath{\rvecv}{\bimV}
\safemath{\rvecw}{\bimW}
\safemath{\rvecx}{\bimX}
\safemath{\rvecy}{\bimY}
\safemath{\rvecz}{\bimZ}

\safemath{\rvecxi}{\bmxi}
\safemath{\rveclambda}{\bmlambda}
\safemath{\rvecmu}{\bmmu}
\safemath{\rvectheta}{\bmtheta}
\safemath{\rvecphi}{\bmphi}

\safemath{\rmatA}{\bimA}
\safemath{\rmatB}{\bimB}
\safemath{\rmatC}{\bimC}
\safemath{\rmatD}{\bimD}
\safemath{\rmatE}{\bimE}
\safemath{\rmatF}{\bimF}
\safemath{\rmatG}{\bimG}
\safemath{\rmatH}{\bimH}
\safemath{\rmatI}{\bimI}
\safemath{\rmatJ}{\bimJ}
\safemath{\rmatK}{\bimK}
\safemath{\rmatL}{\bimL}
\safemath{\rmatM}{\bimM}
\safemath{\rmatN}{\bimN}
\safemath{\rmatO}{\bimO}
\safemath{\rmatP}{\bimP}
\safemath{\rmatQ}{\bimQ}
\safemath{\rmatR}{\bimR}
\safemath{\rmatS}{\bimS}
\safemath{\rmatT}{\bimT}
\safemath{\rmatU}{\bimU}
\safemath{\rmatV}{\bimV}
\safemath{\rmatW}{\bimW}
\safemath{\rmatX}{\bimX}
\safemath{\rmatY}{\bimY}
\safemath{\rmatZ}{\bimZ}

\safemath{\rmatDelta}{\bimDelta}
\safemath{\rmatLambda}{\bimLambda}
\safemath{\rmatPhi}{\bimPhi}
\safemath{\rmatSigma}{\bimSigma}
\safemath{\rmatOmega}{\bimOmega}
\safemath{\rmatTheta}{\bimTheta}

\usepackage{amssymb}
\usepackage{amsfonts}
\usepackage{mathrsfs}
\usepackage{xspace}
\usepackage{bm}
\usepackage{fancyref}
\usepackage{textcomp}

\usepackage{multirow}
\usepackage{stmaryrd}

\newenvironment{textbmatrix}{	\setlength{\arraycolsep}{2.5pt}%
								\big[\begin{matrix}}{\end{matrix}\big]%
								\raisebox{0.08ex}{\vphantom{M}}}

\def\be{\begin{equation}}
\def\ee{\end{equation}}
\def\een{\nonumber \end{equation}}
\def\mat{\begin{bmatrix}}
\def\emat{\end{bmatrix}}
\def\btm{\begin{textbmatrix}}
\def\etm{\end{textbmatrix}}

\def\ba#1\ea{\begin{align}#1\end{align}}
\def\bas#1\eas{\begin{align*}#1\end{align*}}
\def\bs#1\es{\begin{split}#1\end{split}}
\def\bg#1\eg{\begin{gather}#1\end{gather}}
\def\bml#1\eml{\begin{multline}#1\end{multline}}
\def\bi#1\ei{\begin{itemize}#1\end{itemize}}

\newcommand{\lefto}{\mathopen{}\left}

\DeclareMathOperator{\sign}{sign}			
\DeclareMathOperator*{\argmin}{arg\;min}		
\DeclareMathOperator*{\argmax}{arg\;max}		
\DeclareMathOperator{\Exop}{\opE}			
\DeclareMathOperator{\Varop}{\opV\!\mathrm{ar}} 

\newcommand{\abs}[1]{\lefto\lvert#1\right\rvert}		

\newcommand{\vecnorm}[1]{\lefto\lVert#1\right\rVert}		

\safemath{\dirac}{\delta}					
\safemath{\krond}{\dirac}					

\safemath{\upto}{\uparrow}
\safemath{\downto}{\downarrow}
\safemath{\iu}{j}							
\safemath{\ev}{\lambda}						
\safemath{\hilseqspace}{l^{2}}				
\newcommand{\banachfunspace}[1]{\setL^{#1}}	
\safemath{\hilfunspace}{\banachfunspace{2}}	

\safemath{\SNR}{\textit{SNR}} 				
\safemath{\PAR}{\textit{PAR}} 				
\safemath{\No}{N_0}							
\safemath{\Es}{E_s}							
\safemath{\Eb}{E_b}							
\safemath{\EbNo}{\frac{\Eb}{\No}}
\safemath{\EsNo}{\frac{\Es}{\No}}

\DeclareMathOperator{\CHop}{\ensuremath{\opH}} 
\safemath{\tvir}{\rndh_{\CHop}}				
\safemath{\tvtf}{\rndl_{\CHop}}				
\safemath{\spf}{\rnds_{\CHop}}				
\safemath{\bff}{H_{\CHop}}					

\safemath{\ircf}{r_{h}}						
\safemath{\tftvcf}{r_{s}}					
\safemath{\tfcf}{r_{l}}						
\safemath{\bfcf}{r_{H}}						

\safemath{\tcorr}{c_h}						
\safemath{\scf}{c_{s}}						
\safemath{\tfcorr}{c_{l}}					
\safemath{\fcorr}{c_{H}}						

\safemath{\mi}{I}							
\safemath{\capacity}{C}						

\safemath{\normal}{\mathcal{N}}			
\safemath{\jpg}{\mathcal{CN}}			
\safemath{\mchain}{\leftrightarrow}		

\safemath{\dB}{\,\mathrm{dB}}
\safemath{\dBm}{\,\mathrm{dBm}}
\safemath{\Hz}{\,\mathrm{Hz}}
\safemath{\kHz}{\,\mathrm{kHz}}
\safemath{\MHz}{\,\mathrm{MHz}}
\safemath{\GHz}{\,\mathrm{GHz}}
\safemath{\s}{\,\mathrm{s}}
\safemath{\ms}{\,\mathrm{ms}}
\safemath{\mus}{\,\mathrm{\text{\textmu}s}}
\safemath{\ns}{\,\mathrm{ns}}
\safemath{\ps}{\,\mathrm{ps}}
\safemath{\meter}{\,\mathrm{m}}
\safemath{\mm}{\,\mathrm{mm}}
\safemath{\cm}{\,\mathrm{cm}}
\safemath{\m}{\,\mathrm{m}}
\safemath{\W}{\,\mathrm{W}}
\safemath{\mW}{\, \mathrm{mW}}
\safemath{\J}{\,\mathrm{J}}
\safemath{\K}{\,\mathrm{K}}
\safemath{\bit}{\,\mathrm{bit}}
\safemath{\nat}{\,\mathrm{nat}}

\safemath{\define}{\triangleq}			

\safemath{\equivalent}{\sim}
\safemath{\distas}{\sim}					
\safemath{\sdiff}{\Delta}				

\safemath{\reals}{\mathbb{R}}
\safemath{\positivereals}{\reals_{+}}
\safemath{\integers}{\mathbb{Z}}
\safemath{\posint}{\integers_{+}}
\safemath{\naturals}{\mathbb{N}}
\safemath{\posnaturals}{\naturals_{+}}
\safemath{\complexset}{\mathbb{C}}
\safemath{\rationals}{\mathbb{Q}}

\newcommand*{\fancyrefapplabelprefix}{app}		
\newcommand*{\fancyrefthmlabelprefix}{thm}		
\newcommand*{\fancyreflemlabelprefix}{lem}		
\newcommand*{\fancyrefcorlabelprefix}{cor}		
\newcommand*{\fancyrefdeflabelprefix}{def}		
\newcommand*{\fancyrefproplabelprefix}{prop}		
\newcommand*{\fancyrefexmpllabelprefix}{exmpl}
\newcommand*{\fancyrefalglabelprefix}{alg}      
\newcommand*{\fancyreftbllabelprefix}{tbl}		

\frefformat{vario}{\fancyrefseclabelprefix}{Section~#1}
\frefformat{vario}{\fancyrefthmlabelprefix}{Theorem~#1}
\frefformat{vario}{\fancyreftbllabelprefix}{Table~#1}
\frefformat{vario}{\fancyreflemlabelprefix}{Lemma~#1}
\frefformat{vario}{\fancyrefcorlabelprefix}{Corollary~#1}
\frefformat{vario}{\fancyrefdeflabelprefix}{Definition~#1}
\frefformat{vario}{\fancyreffiglabelprefix}{Fig.~#1}
\frefformat{vario}{\fancyrefapplabelprefix}{Appendix~#1}
\frefformat{vario}{\fancyrefeqlabelprefix}{(#1)}
\frefformat{vario}{\fancyrefproplabelprefix}{Proposition~#1}
\frefformat{vario}{\fancyrefexmpllabelprefix}{Example~#1}
\frefformat{vario}{\fancyrefalglabelprefix}{Algorithm~#1}
 \newtheorem{thm}{Theorem}
 \newtheorem{cor}[thm]{Corollary}   
 
 \newtheorem{defi}{Definition}
 \newtheorem{lem}[thm]{Lemma}
 
\newtheorem{asm}{Assumption}

\safemath{\dictab}{[\,\dicta\,\,\dictb\,]}

\safemath{\ysig}{\bmy}
\safemath{\ysighat}{\hat{\ysig}}
\safemath{\ysigdim}{M}
\safemath{\xsig}{\bmx}
\safemath{\xsigdim}{N}
\safemath{\nx}{n_x}
\safemath{\zsig}{\bmz}
\safemath{\zsigdim}{\ysigdim}
\safemath{\rsig}{\bmr}
\safemath{\Adict}{\bA}
\safemath{\Adicttilde}{\widetilde{\Adict}}
\safemath{\Adictdim}{\outputdim\times\xsigdim}
\safemath{\avec}{\bma}
\safemath{\avectilde}{\tilde{\avec}}
\safemath{\Bdict}{\bB}
\safemath{\Bdicttilde}{\widetilde{\Bdict}}
\safemath{\Cdict}{\bC}
\safemath{\cvec}{\bmc}
\safemath{\Ddict}{\bD}
\safemath{\Ddictdim}{\ysigdim\times\xsigdim}
\safemath{\dvec}{\bmd}
\safemath{\Ddicttilde}{\widetilde{\bD}}
\safemath{\Bonb}{\bB}
\safemath{\bvec}{\bmb}
\safemath{\Bonbdim}{\ysigdim\times\ysigdim}
\safemath{\noise}{\bmn}
\safemath{\noisedim}{\ysigim}
\safemath{\err}{\bme}
\safemath{\errdim}{\ysigdim}
\safemath{\errset}{\setE}
\safemath{\nerr}{n_e}
\safemath{\delop}{\bP_\errset}
\safemath{\delopc}{\bP_{{\errset}^c}}

\safemath{\cplxi}{\imath}
\safemath{\cplxj}{\jmath}

\safemath{\dict}{\matD}
\safemath{\inputdim}{N}		
\safemath{\outputdim}{M}		
\safemath{\sparsity}{S}	
\safemath{\inputdimA}{{N_a}}	
\safemath{\inputdimB}{{N_b}}	
\safemath{\elemA}{{n_a}}	
\safemath{\elemB}{{n_b}}	
\safemath{\resA}{\matR_a}	
\safemath{\resB}{\matR_b}	
\safemath{\subD}{\matS} 
\safemath{\subA}{\matS_a} 
\safemath{\subB}{\matS_b} 
\safemath{\dicta}{\matA} 	
\safemath{\dictb}{\matB} 	
\safemath{\hollowS}{H}
\safemath{\hollowA}{H_a}
\safemath{\hollowB}{H_b}
\safemath{\cross}{Z}
\safemath{\coh}{\mu_d}			
\safemath{\coha}{\mu_a}			
\safemath{\cohb}{\mu_b}			
\safemath{\mubs}{\nu}	
\safemath{\cohm}{\mu_m} 
\safemath{\dictset}{\setD}	
\safemath{\dictsetp}{\dictset(\coh,\coha,\cohb)}	
\safemath{\dictsetgen}{\dictset_\text{gen}}
\safemath{\dictsetgenp}{\dictsetgen(\coh)}
\safemath{\dictsetonb}{\dictset_\text{onb}}
\safemath{\dictsetonbp}{\dictsetonb(\coh)}

\safemath{\leftside}{U}
\safemath{\rightsideA}{R_a}
\safemath{\rightsideB}{R_b}

\safemath{\indexS}{\setI_S} 

\safemath{\na}{n_a}			
\safemath{\nb}{n_b}			
\safemath{\coeffa}{p_i}	
\safemath{\coeffb}{q_j}	
\safemath{\seta}{\setP}		
\safemath{\setb}{\setQ}     
\safemath{\setw}{\setW}	
\safemath{\setz}{\setZ}	
\safemath{\cola}{\veca}		
\safemath{\colb}{\vecb}		
\safemath{\cold}{\vecd}		
\safemath{\inputvec}{\vecx} 	
\safemath{\error}{\vece}	
\safemath{\noiseout}{\vecz} 	
\safemath{\inputvecel}{x}
\safemath{\inputveca}{\vecx_a}
\safemath{\inputvecb}{\vecx_b}
\safemath{\outputvec}{\vecy}	
\safemath{\lambdamin}{\lambda_{\mathrm{min}}}

\safemath{\elltwo}{\ell_2}
\safemath{\ellone}{\ell_1}
\safemath{\ellzero}{\ell_0}
\safemath{\ellinf}{\ell_\infty}
\safemath{\ellinftilde}{\ell_{\widetilde\infty}}
\safemath{\licard}{Z(\coh,\coha,\cohb)}
\safemath{\xsol}{\hat{x}}
\safemath{\xbord}{x_b}		
\safemath{\xstat}{x_s}		
\safemath{\xstatLone}{\tilde{x}_s}
\safemath{\order}{\mathcal{O}} 
\safemath{\scales}{\Theta} 
\safemath{\ones}{\mathbf{1}} 
\safemath{\zeroes}{\mathbf{0}} 
\safemath{\thlone}{\kappa(\coh,\cohb)} 
\safemath{\constoneA}{\delta} 
\safemath{\constoneB}{\epsilon} 
\safemath{\nlarge}{L}				   
\safemath{\sumlarge}{S_\nlarge}
\safemath{\maxlarger}{P_\nlarge}	   
\safemath{\Pzero}{\textrm{P0}}
\safemath{\Pone}{\textrm{P1}}
\safemath{\vecfir}{\vecw}			 
\safemath{\vecsec}{\vecz}
\safemath{\elvecfir}{w}              
\safemath{\elvecsec}{z}				 
\safemath{\nlargefir}{n}
\safemath{\normout}{\gamma}
\safemath{\auxfun}{h}
\safemath{\supp}{\textrm{supp}}

\safemath{\indexa}{\ell}
\safemath{\indexb}{r}
\safemath{\indexc}{i}
\safemath{\indexd}{j}

\safemath{\project}{P}

\newcommand{\Sr}{S_\textnormal{R}}%
\newcommand{\dd}{\textnormal{d}}%

\usepackage{color}

\newcommand{\sellr}[1]{r^\text{R}#1}
\newcommand{\selli}[1]{r^\text{I}#1}

\newcommand{\revision}[1]{\textcolor{black}{#1}}
\newcommand{\revisiontwo}[1]{\textcolor{black}{#1}}
\newcommand{\revisionthree}[1]{\textcolor{black}{#1}}

\newcommand{\realpart}[1]{\textnormal{Re}\!\left\{ #1 \right\}\!}

\newcommand{\imagpart}[1]{\textnormal{Im}\!\left\{ #1 \right\}\!}

\safemath{\LAMA}{\textrm{LAMA}}

\safemath{\mLAMA}{\textrm{M-LAMA}}
\safemath{\smLAMA}{\textrm{SM-LAMA}}
\safemath{\mCBAMP}{\textrm{mcB-AMP}}

\safemath{\MRT}{\textrm{MRT}}
\safemath{\betamax}{\beta^\textnormal{max}}
\safemath{\tmax}{t_\textnormal{max}}
\safemath{\betamaxno}{\beta^\textnormal{max}}
\safemath{\betamin}{\beta^\textnormal{min}}
\safemath{\betaminno}{\beta^\textnormal{min}}

\safemath{\Nomin}{\No^\textnormal{min}(\beta)}
\safemath{\Nominnobeta}{\No^\textnormal{min}}
\safemath{\Nomax}{\No^\textnormal{max}(\beta)}
\safemath{\Nomaxnobeta}{\No^\textnormal{max}}

\safemath{\imageTxt}{{\bf \textnormal{i}}}

\safemath{\MAP}{\textrm{MAP}}
\safemath{\IO}{\textrm{IO}}
\safemath{\JO}{\textrm{JO}}
\safemath{\Nopost}{N_{0}^\textnormal{post}}
\safemath{\MT}{{M_\textnormal{T}}}
\safemath{\MR}{{M_\textnormal{R}}}
\safemath{\Tran}{\textnormal{T}}
\safemath{\Herm}{\textnormal{H}}
\safemath{\row}{\textnormal{r}}
\safemath{\col}{\textnormal{c}}

\safemath{\gammasq}{\revision{\tau}}
\safemath{\gammapaper}{\revision{\tau}}

\safemath{\psimm}{\revision{\Psi^\textnormal{mm}}}
\safemath{\psimmstar}{\revision{\Psi^\textnormal{mm}_\star}}
\begin{document}

\title{Mismatched Data Detection in Massive MU-MIMO}

\author{Charles Jeon, Arian Maleki, and Christoph Studer
\thanks{Parts of this paper were presented at the IEEE International Symposium on Information Theory (ISIT)~\cite{JMS2016}.}
\thanks{C. Jeon was with the School of Electrical and Computer Engineering, Cornell University, Ithaca, NY, and is now with Apple Inc., San Diego, CA; e-mail: {cj339@cornell.edu}}
\thanks{A. Maleki is with Department of Statistics at Columbia University, New York City, NY; e-mail: {arian@stat.columbia.edu}.}
\thanks{\revision{C.~Studer is with the Department of Information Technology and Electrical Engineering, ETH Z\"urich, Z\"urich, Switzerland; e-mail: {studer@ethz.ch}}}
}

\maketitle

\begin{abstract}
We investigate mismatched data detection for massive multi-user (MU) multiple-input multiple-output (MIMO) wireless systems in which the prior distribution of the transmit signal used in the data detector differs from the true prior.
In order to minimize the performance loss caused by the prior mismatch, we include a tuning stage into the recently proposed large-MIMO approximate message passing (LAMA) algorithm, which enables the development of data detectors with optimal as well as sub-optimal parameter tuning.
\revision{We show that carefully-selected priors enable the design of simpler and computationally more efficient data detection algorithms compared to LAMA that uses the optimal prior, while achieving near-optimal error-rate performance.}
In particular, we demonstrate that a hardware-friendly approximation of the exact prior enables the design of low-complexity data detectors that achieve near individually-optimal performance.
Furthermore, for Gaussian priors and uniform priors within a hypercube covering the quadrature amplitude modulation (QAM) constellation, our performance analysis recovers classical and recent results on linear and non-linear massive MU-MIMO data detection, respectively.
\end{abstract}



\section{Introduction}
\label{sec:intro}

Data detection in multiple-input multiple-output (MIMO) systems deals with the recovery of the data vector $\bms_0\in\setO^\MT$, where $\setO$ is a finite constellation (e.g., QAM or PSK), from the noisy input-output relation \mbox{$\vecy=\bH\vecs_0+\bmn$}. In what follows, $\MT$ and $\MR$ denotes the number of transmit and receive antennas, respectively, $\bmy\in\complexset^\MR$ is the receive vector, \mbox{$\bH\in\complexset^{\MR\times\MT}$} is the known MIMO system matrix, and \mbox{$\vecn\in\complexset^\MR$} is  i.i.d.\ circularly symmetric complex Gaussian noise \revision{with variance~$\No$}.  \revision{In order} to minimize the symbol error rate, we are interested in solving the following individually-optimal (IO) data detection problem \cite{V1998,guo2003multiuser,GV2005}:
\begin{align*}
(\text{IO})\quad
s_\ell^\text{IO} & = \argmax_{\tilde s_\ell\in\setO} \,p\!\left(\tilde s_\ell \,|\, \bmy, \bH\right)\!, \,\,\, \ell=1,\ldots,\MT.
\end{align*}
Here,  $s_\ell^\text{IO}$ denotes the $\ell$th IO estimate and $p\!\left(\tilde s_\ell \,|\, \bmy, \bH\right)$ is the conditional probability density function of $\tilde s_\ell^\text{IO}$ given the receive vector and the channel matrix.

Since the IO data detection problem is of combinatorial nature~\cite{V1998,guo2003multiuser,GV2005}, an exhaustive search or sphere-decoding methods~\cite{SJSB11} would result in prohibitive  complexity for systems where~$\MT$ is large.
To alleviate this complexity bottleneck, the algorithm proposed in~\cite{JGMS2015conf}, referred to as large MIMO approximate message passing (LAMA), achieves the error-rate performance of the IO data detector using a simple iterative procedure in the large-system limit, i.e., for i.i.d.\ Gaussian channel matrices with a fixed system ratio $\beta=\MT/\MR$ and $\MT\to\infty$.
Although the theoretical performance guarantees for LAMA only hold in the large-system limit, the algorithm delivers near-IO performance for practical (finite-dimensional) systems at low complexity~\cite{JGMS2015conf}.
\revision{Despite all of these advantages, LAMA requires repeated computations of transcendental functions (e.g., exponentials) at excessively high arithmetic precision, which render the design of high-throughput hardware implementations that rely on finite precision (fixed-point) arithmetic a challenging task.}

\subsection{Contributions}
\revision{In order to address these hardware-limitation aspects,} we develop a mismatched version of the complex Bayesian approximate message passing (cB-AMP) framework proposed in \cite{JGMS2015conf} that includes a tuning stage to minimize the performance loss caused by a mismatch in the signal prior.
To enable a precise performance analysis in the large-system limit, we develop a mismatched state-evolution (SE) framework.
The proposed framework enables the design of new data detection algorithms and their exact performance analysis in the large-system limit.
Our key contributions are as follows.
\begin{itemize}
\item We propose a mismatched version of the LAMA algorithm~\cite{JGMS2015conf} (short \mLAMA{}), which allows carefully-selected mismatched priors that enable near-IO performance \revision{while avoiding the computation of transcendental functions and relaxing numerical precision requirements.}
\item We show that \mLAMA{} is a generalization of LAMA \cite{JGMS2015conf} by proving that \mLAMA reduces to LAMA when there is no mismatch in the prior distribution.
\item We demonstrate that  \mLAMA{} with a Gaussian prior recovers classical results for linear data detectors in \cite{TH1999}.
\item We demonstrate that \mLAMA{} with a uniform prior within a hypercube  covering PAM/QAM constellations recovers recent results from convex-optimization-based data detection methods analyzed in \cite{TAXH2015,TXH2018,ASH2019,HL2020}.%
\item We demonstrate that a novel Gray-coding based approximation for PAM/QAM constellations enables near-IO performance at significantly reduced complexity.
The resulting approximation has been implemented recently in a digital integrated circuit prototype \cite{JCS2019}, \revision{which showcases the practicality of our framework}.
\item We provide simulation results in finite-dimensional systems to confirm that the developed theory accurately characterizes the performance of mismatched data detectors even for moderately-sized massive MU-MIMO systems.
\end{itemize}

\subsection{Related Work}

Linear data-detection algorithms for MIMO systems, such as zero forcing (ZF) or minimum mean-square error (MMSE) equalization, are well-known instances of mismatched data detectors. 
The performance of such linear data detectors in the asymptotic large-system limit has been investigated \revisiontwo{in~\cite{VS1999,TH1999,SV2001,TanakaCDMA}}.
Another instance of mismatched data detection is the so-called box-relaxation data detector, which relaxes the discrete constellation to its convex hull \cite{tan2001constrained,yener2002cdma,pan2014mimo,TAXH2015}.
Corresponding theoretical results in \cite{DT2011,MR2011} for noiseless systems show that a system ratio of $\beta<2$ enables perfect signal recovery. The recovery performance for the noisy case was analyzed recently in \cite{TAXH2015,TXH2018,ASH2019,HL2020}.
The framework presented in our paper recovers all of these results while enabling the design of novel, more general, and computationally efficient algorithms.
The mismatched LAMA (\mLAMA) algorithm proposed in this paper relies upon approximate message passing (AMP)~\cite{donoho2009,BM2011,andreaGMCS,BMDK2020,APRCN2020,FJC2019}, which was developed for sparse signal recovery and compressive sensing.
The case of mismatched estimation of sparse signals via AMP was first studied in~\cite{MMB2013}, where the performance of AMP was analyzed when the true prior is unknown.
The AMP algorithm in \cite{MMB2013,MMB2015} includes a tuning stage for optimal parameter selection, which minimizes the output mean-squared error (MSE).
The key differences between the results in \cite{MMB2013,MMB2015} and \mLAMA{} as proposed here are that (i) we consider data detection in MU-MIMO systems and (ii) we know the true signal prior but intentionally select a mismatched prior in order to design hardware-friendly data detection algorithms that enable near-IO performance.

\subsection{Notation}
Lowercase and uppercase boldface letters stand for vectors and matrices, respectively. We define the adjoint of a matrix~$\bH$ as $\bH^\Herm$.
We use $\left\langle\cdot\right\rangle$ to abbreviate $\left\langle \bmx \right\rangle = \frac{1}{N}\sum_{k=1}^N x_k$.
A multivariate zero-mean circular symmetric complex-valued Gaussian probability density function (PDF) is denoted by $\setC\setN( \bm{0} ,\bK)$, where~$\bK$ the covariance matrix. $\Exop_X\!\left[\cdot\right]$ and $\Varop_X\!\left[\cdot\right]$ denotes the expectation and variance operator with respect to the PDF of the random variable~$X$, respectively.
We define $\Phi(x)$ and $Q(x)$ as the cumulative density function (CDF) and $Q$-function for a standard real-valued Gaussian, i.e., $\Phi(x) = \int_{-\infty}^x e^{-t^2/2}/\sqrt{2\pi} \dd t$ and $Q(x) = 1-\Phi(x)$.


\section{Mismatched Complex Bayesian AMP}\label{sec:cB-AMP}

We start by presenting a mismatched version of the complex Bayesian approximate message passing (cB-AMP) framework in~\cite{JGMS2015} (short \mCBAMP), which enables the use of different prior distributions $\tilde p(\cdot)$ than the true signal prior $p(\cdot)$.
To minimize the performance loss caused by the mismatched prior, we include a tuning stage into the  \mCBAMP framework.

\subsection{The \mCBAMP Framework}\label{sec:MCBAMP_alg}

Given an i.i.d. prior distribution $p(\bms_0)=\prod_{\ell=1}^N  p(s_{0\ell})$ of the true signal $\bms_0$ and a mismatched prior distribution $\tilde  p(\tilde \bms)=\prod_{\ell=1}^N  \tilde p(\tilde s_\ell)$,
the mismatched cB-AMP algorithm corresponds to the following iterative procedure:
\begin{align}
\label{eq:decouple_estimate}
\tilde\sigma^2_{t} &= \textstyle \frac{1}{\MR}\vecnorm{\bmr^{t}}_2^2,\\
\label{eq:tau_opt}
\revision{
\tilde\tau^{t}} &= \revision{\argmin_{\tau \geq0} \Psi^{\text{mm}}(\tilde\sigma_t^2, \tau ),}
\\
\label{eq:F_tau}
\bms^{t+1} &= \mathsf{F}^\text{mm}\!\left(\bms^{t} + \bH^\Herm \bmr^{t},\revision{
\tilde\tau^{t}}\right)\!,\\
\label{eq:Req}
\bmr^{t+1} &= \bmy - \bH\bms^{t+1} + \beta \bmr^t \!\left\langle\mathsf{F'}^\text{mm}(\bms^t+\bH^\Herm\bmr^t,\revision{
\tilde\tau^{t}})
\right\rangle\!,
\end{align}
which is carried out for $t_\text{max}$ iterations $t=1,\ldots,t_\text{max}$.
The algorithm is initialized by $s^1_\ell = \Exop_{S_0}[S_0]$ for all $\ell=1,\ldots,\MT$, where $S_0$ is a random variable distributed as $S_0\sim p(s_0)$, $\bmr^1 = \bmy - \bH\bms^1$, and $\mathsf{F'}^\text{mm}$ is the derivative of $\mathsf{F}^\text{mm}$ in the first argument.
The functions $\mathsf{F}^\text{mm}$ and $\mathsf{F'}^\text{mm}$ operate element-wise on vectors (in their first argument).
\revision{
In \fref{eq:tau_opt}, the variance parameter~$\tilde\tau^t$ is selected to minimize the mean-squared error when using the mismatched prior defined by
\begin{align}
\label{eq:PSI_mm}
\Psi^{\text{mm}}(\sigma^2, \tau ) = \Exop_{S_0,Z}\!\left[\abs{\mathsf{F}^\text{mm}(S_0+\sigma Z,\tau)-S_0}^2\right]\!.
\end{align}
Here, expectation is taken with respect to the true prior distribution of $S_0$ and $Z\sim\setC\setN(0,1)$.
}

In what follows, the function $\mathsf{F}^\text{mm}(s_\ell,\revision{
\tilde\tau})$ is the posterior mean \revision{with respect to} the mismatched prior distribution $\tilde p(\tilde s_\ell)$ and variance parameter $\revision{
\tilde\tau}$ that is given by
\begin{align}\label{eq:F}
\mathsf{F}^\text{mm}(s_\ell,\revision{
\tilde\tau})
&= \Exop_{\tilde S}[\tilde S\vert s_\ell]
= \int_\complexset \tilde s p(\tilde s\vert s_\ell,\revision{
\tilde\tau})\dd \tilde s.
\end{align}
Here, $p(\tilde s\vert s_\ell,\revision{
\tilde\tau})$ is the posterior PDF defined as $p(\tilde s\vert s_\ell,\revision{
\tilde\tau})=\frac{1}{Z}p(s_\ell\vert\tilde s,\revision{
\tilde\tau}) \tilde p(\tilde s)$ with $p( s_\ell\vert\tilde s,\revision{
\tilde\tau})\sim\setC\setN(\tilde s,\revision{
\tilde\tau})$ and  $Z$ is a normalization constant.

\revision{We emphasize that the \mCBAMP algorithm differs from the original cB-AMP algorithm in \cite{JGMS2015} by the additional steps in~\fref{eq:decouple_estimate} and \fref{eq:tau_opt}.}
At every iteration, step \fref{eq:decouple_estimate} estimates the decoupled noise variance $\sigma^2_t$
\revision{(see \fref{sec:optimal_tuning} for a discussion)} and step~\fref{eq:tau_opt} optimally tunes the variance parameter $\tau^t$ based on the estimate for $\sigma^2_t$.
The tuning stage in \fref{eq:tau_opt} ensures that the \mCBAMP algorithm converges to the solution that minimizes the so-called decoupled noise variance $\sigma^2_t$ in every algorithm iteration; see \fref{sec:optimal_tuning} for the proof.

\subsection{Decoupling Property of AMP-based Algorithms}\label{sec:decoupling_section}
\revision{In what follows, we make the following assumption.
\begin{asm}
For a system with $\MT$ transmit antennas and~$\MR$ receive antennas, we assume entries of the $\MR\times\MT$ channel matrix~$\bH$ be i.i.d.\ circularly symmetric complex Gaussian (CSCG) with variance $1/\MR$.
\end{asm}
For the subsequent analysis, we use the following definition.}
\begin{defi}
We define the \emph{large-system limit} by fixing the system ratio $\beta = \MT/\MR$ and by letting $\MT\to\infty$. \label{def:largesystemlimit}
\end{defi}

As shown in \cite{JGMS2015,Maleki2010phd,BM2011}, AMP-based algorithms effectively decouple the MIMO system into parallel AWGN channels in the large-system limit, i.e., the quantity $\bms^t+\bH^\Herm\bmr^t$ can be expressed as $\bms_0+ \bmw^t$, where $\bmw^t\sim\setC\setN(0,\sigma_t^2\bI_\MT)$ and~$\sigma_t^2$ is the decoupled noise variance.
A key property of AMP-based algorithms is that the decoupled noise variance $\sigma_t^2$ can be tracked \emph{exactly} by the state evolution (SE) framework.
The mismatched SE framework for \mCBAMP algorithm is detailed in \fref{thm:SE}.
We note that the mismatched SE framework is an instance of the SE framework analyzed in \cite{BM2011}, where we use the posterior mean function derived from the mismatched prior in \fref{eq:F}.
\revisiontwo{The proof follows from \cite[Thm.~1]{BM2011} (see \cite[Sec.~3.4]{BM2011} for the proof), where we define the function  $\psi$ as follows: $\psi(x_i^{t+1},x_{0,i}) = \abs{x_i^{t+1} -x_{0,i}}$.
}

\begin{thm}[\revisionthree{\!\!\cite[Thm.~1]{BM2011}}] \label{thm:SE}
Suppose that $p(\bms_0)=\prod_{\ell=1}^\MT p(s_{0\ell})$ and $\tilde p(\tilde\bms)=\prod_{\ell=1}^\MT \tilde p(\tilde s_{\ell})$. Assume the large-system limit and that $\mathsf{F}^\textnormal{mm}$ is a \revision{Lipschitz-continuous function}.
Then, the decoupled noise variance $\sigma_{t+1}^2$ after $t$ iterations of \mCBAMP is given by the following coupled recursion:
\begin{align}
\label{eq:SE_MSE}
\sigma_{t+1}^2 &= \No + \beta\Psi^\textnormal{mm}(\sigma_t^2,\gammasq^t ),
\end{align}
which is initialized by $\sigma^2_1 = \No + \beta\Varop_{S_0}[S_0]$. Here, $S_0\sim p(s_0)$  and the MSE function is defined \revision{
   \fref{eq:PSI_mm}.
\revision{Here, $\tau^t$ is a tuning parameter that can, in principle, be chosen arbitrarily in each iteration $t$.}
}
\end{thm}

\revisiontwo{
We note that SE framework from \cite{BM2011} is valid for any Lipschitz-continuous function $\mathsf{F}^\text{mm}$ with a fixed $\tau^t$. Since the function $\mathsf{F}^\text{mm}$ is dependent on its second argument $\tau^t$ via \fref{eq:F}, the choice of $\tau^t$ at each iteration $t$ influences the SE recursion in \fref{eq:SE_MSE}.
}
\revision{We now move on to discussing how to properly select the tuning parameter $\tau^t$ at every iteration.}

\subsection{Optimal Tuning of the Variance Parameter $\tau$}\label{sec:optimal_tuning}
The purpose of the tuning stage in \fref{eq:tau_opt} is to optimally set the variance parameter~$\tau^t$ in every iteration $t$, which is used to compute the posterior mean in \fref{eq:F}.
Before we discuss the tuning procedure in detail, we define what we mean by optimally-tuning the variance parameter $\tau^t$.
For $t=1,\ldots,t_\text{max}$ iterations, our goal is to minimize the decoupled noise variance $\sigma_{t_\text{max}+1}^2$ given by \fref{thm:SE} as the smallest $\sigma_{t_\text{max}+1}^2$ that minimizes the MSE of our algorithm.
To achieve this goal, the optimal choice is to tune the parameters $\{\tau^1,\ldots,\tau^{t_\text{max}}\}$ so that \mCBAMP ultimately leads to the smallest $\sigma_{t_\text{max}+1}^2$.
We next show that the tuning stage \fref{eq:tau_opt}, which is carried out separately at every iteration, in fact achieves the smallest $\sigma_{t_\text{max}+1}^2$, i.e., optimally tunes the variance parameters~$\tau^t$, i.e.,
\revision{
\begin{align}
\label{eq:SE_gamma}
\gammasq^t &= \argmin_{\gammasq\geq0}\Psi^\textnormal{mm}(\sigma_t^2,\gammasq).
\end{align}
}
We note that suboptimal choices of $\tau^t$ can either lead to a higher $\sigma_{t_\text{max}+1}^2$ or \revision{converge more slowly to} the minimal $\sigma_{t_\text{max}+1}^2$.
%
In addition, if the true prior is identical to the mismatched prior, i.e., $p(\bms_0)=\tilde p(\tilde \bms)$, then cB-AMP in \cite{JGMS2015conf} selects optimally-tuned parameters according to \fref{eq:tau_opt}.
Therefore, \mCBAMP results in the same  decoupled noise variance as that given by cB-AMP.  The proof of the following result is given in \fref{app:lama_already_opt}.
\begin{lem} If there is no prior mismatch, i.e., $p(\bms_0)=\tilde p(\tilde \bms)$, then the decoupled noise variance $\sigma^2_{t+1}$ of \mCBAMP is equivalent to $\sigma^2_{t+1}$ of cB-AMP \cite[Eq.~3]{JGMS2015conf}.\label{lem:lama_already_opt}
\end{lem}

We now define ``optimally-tuned'' parameters using the definition of \cite{MMB2015}.

\begin{defi}\label{def:asymp_opt} Assume the large-system limit and denote the decoupled noise variance of \mCBAMP obtained  from the sequence $\{\tau^1,\ldots,\tau^{t_\text{max}}\}$ as $\sigma^2_{t_\text{max}+1}(\tau^1,\ldots,\tau^{t_\text{max}})$.
A sequence of parameters $\{\tau^1_\star,\ldots,\tau^{t_\text{max}}_\star\}$ is optimally-tuned at the iteration $t_\text{max}$, if and only if for all $\{\tau^1,\ldots,\tau^{t_\text{max}}\} \in [0,\infty)^{t_\text{max}}$,
\begin{align}\label{eq:optimal_tune_MSE}
\sigma^2_{t_\text{max}+1}(\tau^1_\star,\ldots,\tau^{t_\text{max}}_\star)
\leq
\sigma^2_{t_\text{max}+1}(\tau^1,\ldots,\tau^{t_\text{max}}).
\end{align}
\end{defi}

In words, a sequence of optimally-tuned parameters minimizes the decoupled noise variance $\sigma_{t_\text{max}+1}^2$ defined in \fref{thm:SE} given by \mCBAMP after $t_\text{max}$ iterations.
\revision{
We note that the sequences $\{\tau^1_\star,\ldots,\tau^t_\star\}$  are computed recursively for each iteration $t$ so that different $\{\tau^1,\ldots,\tau^t\}$ sequences will lead to different values of $\sigma^2_t$.
Since our theoretical results are primarily for optimal tuning, we will drop the $\{\tau^1_\star,\ldots,\tau^{t}_\star\}$ sequences in $\sigma_{t}^2 (\tau_\star^1,\ldots,\tau_\star^t)$, and use $\sigma_{t}^2$ throughout this paper.
}

The following theorem shows that \mCBAMP leads to the $t_\text{max}$ optimally-tuned parameters $\{\tau^1_\star,\ldots,\tau^{t_\text{max}}_\star\}$.
For the sake of brevity, we skip the proof details as it closely follows the proof given in~\cite[Sec. 4.4]{MMB2015} with minor modifications.

\begin{thm}\label{thm:thm3_7_mbb} \revision{\cite[{Thm.~3.7}]{MMB2015}} Suppose $\{\tau_\star^1,\ldots,\tau_\star^{t_\text{max}}\}$ are optimally-tuned for iteration $t_\text{max}$. Then, for any $t<t_\text{max}$, the parameters $\{\tau_\star^1,\ldots,\tau_\star^{t}\}$ are also optimally-tuned for iteration $t$.
Thus, one can obtain $t_\text{max}$ optimally-tuned variance parameters by optimizing $\tau^1_\star$ at $t=1$, and then proceeding iteratively by optimizing $\tau^t_\star$ until $t=t_\text{max}$.
\end{thm}

The exact value of the decoupled noise variance $\sigma_t^2$ that is needed for the tuning stage  in \fref{eq:tau_opt} to select $\tau^t_\star$ is, in general, unknown at iteration $t$.
\revisionthree{
In place of the decoupled noise variance~$\sigma_t^2$, we use the estimate $\tilde\sigma_t^2 = \frac{1}{\MR}\vecnorm{\bmr^t}^2$ in step~\fref{eq:decouple_estimate}, which convergences  the true decoupled noise variance~$\sigma_t^2$ in the large-system limit. The following lemma, with proof given in \fref{app:proof_lemma_convergence}, establishes this fact. 
\begin{lem}\label{lem:lem_converge}  For Assumption 1 and the large-system limit, the estimate $\tilde\sigma_t^2 = \frac{1}{\MR}\vecnorm{\bmr^t}^2$ converges to $\sigma_t^2$.
\end{lem}
}
\revisionthree{We note that \fref{lem:lem_converge} holds even when mismatched~priors are used, as long as $\mathsf{F}^\text{mm}$ is a Lipschitz-continuous function.}
\revision{
One consequence of \fref{thm:thm3_7_mbb} and \revisionthree{\fref{lem:lem_converge}} is that the $t_\text{max}$ tuning parameters not only achieve the minimal value of $\sigma^2_{t_\text{max}}(\tau_\star^1,\ldots,\tau_\star^{t_\text{max}})$ with \mCBAMP, but also do so at the fastest convergence rate. This observation is a consequence of the following argument: If $\tau_\star^1$ is optimal for $t=1$, then $\tau_\star^2$ obtained via $\tau_\star^1$ is optimal for $t=2$. We can repeat the same argument until $t_\text{max}$ to arrive at $\tau_\star^{t_\text{max}}$, which implies that $\{\tau_\star\}$ yields the fastest convergence~rate.
}

\subsection{Decomposing Complex-Valued Systems}\label{sec:decomposition}
We now briefly discuss properties of \mCBAMP in complex-valued systems that will be necessary for our analysis of \mCBAMP in massive MU-MIMO systems.
In particular, we show that for certain constellations, the complex-valued set $\setO$ can equivalently be characterized by the real-valued set $\realpart{\setO}$.
\begin{defi}\label{def:sep_const}For all $s\in\setO$, express $s$ as $s = a+\imageTxt b$, where $a \in\realpart{\setO}\,$, $b\in\imagpart{\setO}\,$. Then, the constellation $\setO$ is called \emph{separable} if $p(s) = p(a)p(b)$ holds for all $s\in\setO$ and $\realpart{\setO}=\imagpart{\setO}$.
\end{defi}
An example of a separable constellation is $M^2$-QAM with equally likely transmit symbols. For such a separable set $\setO$, the following lemma (with proof given in \cite{JGMS2015}) shows that
the MSE $\Psi^\textnormal{mm}$ can be equivalently computed from an equivalent MSE from a corresponding real-valued system.

\begin{lem}\label{lem:identical_separable}\revision{\cite[{Lem.~8}]{JGMS2015}}
Let the constellation $\setO$ be separable. Define $\Sr = \realpart{S}$ and denote the real-part of $\setO$ as $\setO^\textnormal{R}$.
Define $\revision{\mathsf{F}_\textnormal{R}}$
as the message mean
function
with $\Sr \sim p(\realpart{S})$. Also define the MSE function $\Psi$ for the real-valued prior $\Sr$ as:
\begin{align}
\revision{\Psi^\textnormal{mm}_\textnormal{R}}(\sigma^2,\gammasq) &= \Exop_{\Sr,Z_\textnormal{R}}\!\left[\!\left(\revision{\mathsf{F}_\textnormal{R}}(\Sr + \sigma Z_\textnormal{R},\gammasq) - \Sr\right)^{\!2}\right]\!,
\end{align}
where $Z_\textnormal{R}\sim\setN(0,1)$.
Then, we have the following relation for $\Psi^\textnormal{mm}$
between the complex-valued constellation $\setO$ and the real-valued constellation $\setO^\textnormal{R}$:
\begin{align}\label{eq:realMSE}
\Psi^\textnormal{mm}(\sigma^2,\gammasq) = 2 \revision{\Psi^\textnormal{mm}_\textnormal{R}}\!\left(\frac{\sigma^2}{2},\frac{\gammasq}{2}\right)\!.
\end{align}
Therefore, the recursions in
\fref{eq:SE_MSE} can be simplified to:
\begin{align}
\sigma_t^2 &= \No + \beta\Psi^\textnormal{mm}(\sigma^2_t,\gammasq^t)
= \No+2\beta\revision{\Psi^\textnormal{mm}_\textnormal{R}}\!\left(\frac{\sigma^2_t}{2},\frac{\gammasq^t}{2}\right)\!.
\end{align}

\end{lem}
\fref{lem:identical_separable} shows that the tuning stage in \fref{eq:tau_opt} and message mean \fref{eq:F_tau} of the mcB-AMP algorithm can be computed (often more efficiently) in parallel for real and imaginary dimensions.

\subsection{Fixed-Point Analysis}\label{sec:fp_betas}

While the performance of \mCBAMP at every iteration $t=1,\ldots,t_\text{max}$ in the large-system limit can be characterized by the SE recursion equations in \fref{thm:SE}, we can analyze the performance of \mCBAMP for $t_\text{max}\to\infty$.
In this case, the mismatched SE in \fref{thm:SE} converges to the following fixed-point equation:
\begin{align}\label{eq:fpequation}
\sigma^2_\star = \No + \beta\min_{\gammasq\geq0}\Psi^\text{mm}(\sigma^2_\star,\gammasq) = \No + \beta \psimmstar(\sigma^2_\star),
\end{align}
\revision{where we defined the minimum mean-square error for the mismatched prior with noise variance $\sigma^2$ as follows:
\begin{align}\label{eq:psiminimum}
\psimmstar ( \sigma^2) = \min_{\tau\geq0} \psimm ( \sigma^2,\tau).
\end{align}
}%
Thus, as $t_\text{max}\to\infty$, the decoupled noise variance by \mCBAMP converges to $\sigma_\star^2$ determined by  \fref{eq:fpequation}.
If there are multiple fixed points, then \mCBAMP,
in general\footnote{\revision{The algorithm may converge to another fixed-point if \mCBAMP is initialized sufficiently close to such a fixed point \cite{ZMWL2015}.}}, converges to the largest fixed-point solution to \fref{eq:fpequation}, which ultimately leads to a higher probability of error than that of the smallest fixed-point solution.
In order to provide conditions on the MIMO system to ensure a unique fixed-point solution to~\fref{eq:fpequation}, we use the following definition from \cite[Def. 2]{JGMS2015conf}.

\begin{defi}
Fix the true prior $p(\bms_0)$ and the mismatched prior $\tilde p(\tilde \bms)$. Then, the minimum recovery threshold (MRT) $\betamin$ is defined by
\begin{align}\label{eq:betamin}
\betamin = \min_{\sigma^2\geq0}
\!\left\{
\!\left(
\frac{\dd \Psi^\text{mm}_\star(\sigma^2)}{\dd \sigma^2}
\right)^{\!-1}
\right\}\!.
\end{align}
\end{defi}
By the definition of $\betamin$, for all system ratios $\beta<\betamin$ regardless of the noise variance $\No$, the fixed-point solution in~\fref{eq:fpequation} is unique.
\revision{
   To see this, we can rewrite the fixed point equation as $\sigma^2 - \beta \Psi(\sigma^2)= N_0$.
   Then, it suffices to show that the function $g(\sigma^2)=\sigma^2 - \beta \Psi(\sigma^2)$ is a strictly increasing sequence in $\sigma^2$.
If $\beta<\beta^{\text{min}}$, then we have:}
\begin{align*}
\revision{\frac{ \text{d} g(\sigma^2) }{ \text{d}\sigma^2 }
=
1 - \beta \frac{ \text{d} \Psi(\sigma^2) }{ \text{d}\sigma^2 } < 1 - \beta^{\text{min}} \frac{ \text{d} \Psi(\sigma^2) }{ \text{d}\sigma^2 } > 0.}
\end{align*}

The following lemma details the existence of a unique fixed-point at $\beta=\betamin$. \revision{\fref{lem:unique_fp_MRT} follows from the state-evolution framework by noting that only one unique value of $\sigma^2$ satisfies the fixed-point equation when $\beta=\betamin$.}
\begin{lem}\label{lem:unique_fp_MRT}
Fix the true prior $p(\bms_0)$ and the mismatched prior~$\tilde p(\tilde \bms)$. Let
\begin{align}\sigma^2_\star=\argmin\limits_{\sigma^2\geq0}
\!\left\{
\!\left(
\frac{\dd \Psi^\text{mm}_\star(\sigma^2)}{\dd \sigma^2}
\right)^{\!-1}
\right\}\!.
\end{align} Then, if for any other $\sigma^2\neq \sigma^2_\star$, \mbox{$
\betamin\dd \Psi^\text{mm}_\star(\sigma^2)/\dd \sigma^2<1$}, \mLAMA has a unique fixed point at $\beta=\betamin$ regardless of the noise variance $\No$.
\end{lem}

In \fref{sec:QAMConstellation}, we will use \fref{lem:unique_fp_MRT} to extract conditions for which \mCBAMP has a unique fixed-point solution for PAM/QAM constellation sets---uniqueness of the fixed-point enables us to precisely characterize the MSE of \mCBAMP{}~\cite{BM2011}.


\section{Mismatched Data Detection with
Optimal
Tuning:
General Case
} \label{sec:GaussianPrior}

We now apply the mismatched cB-AMP framework to mismatched data detection in massive MIMO systems, and refer to the algorithm as  mismatched large MIMO AMP (\mLAMA).
\revision{As noted in \fref{sec:MCBAMP_alg}, \mLAMA{} differs from LAMA by the additional tuning stage. We will first discuss optimal tuning, and then present sub-optimal tuning that allows us to remove the tuning stage completely. }
We start by introducing \mLAMA and then, present the \mLAMA algorithm for a Gaussian priors.

\subsection{Why Should One Use a Mismatched Prior?}

In MIMO systems, the true signal prior is typically known at the receiver. It is therefore natural to ask why the use of a mismatched prior should be useful, especially since the true prior, which leads to the \LAMA algorithm \cite{JGMS2015,JGMS2015conf}, will minimize the probability of error.
To answer this question, we highlight the following practically-relevant advantages of mismatched detectors:
\begin{inparaenum}[(i)]
\item
For the \LAMA algorithm where there is no prior mismatch, the posterior mean function is given by \cite{JGMS2015conf}:
\begin{align}
\label{eq:FrealFunction}
\mathsf{F}(r,\tau) = \frac{\sum\limits_{a\in\setO}a\exp\!\left(-\frac{1}{\tau}\abs{r-a}^2\right)}{\sum\limits_{a\in\setO}\exp\!\left(-\frac{1}{\tau}\abs{r-a}^2\right)}.
\end{align}
Calculating this expression  with double-precision floating point arithmetic becomes numerically unstable for small values of $\tau$.
Hence, the design of high-performance application-specific integrated circuits (ASICs) that deploy finite-precision (fixed-point) arithmetic is extremely difficult. Suitably-chosen mismatched priors can alleviate the need for high arithmetic precision and large dynamic range.
\item While in some situations, the true prior may be unknown to the receiver, some information on the prior may be available (e.g., the energy). We will show in \fref{sec:gaus_prior} that \mCBAMP with a mismatched Gaussian prior enables optimal tuning as in \fref{eq:tau_opt} but only requires knowledge of the energy of the true prior distribution.
\end{inparaenum}

\subsection{Optimally-Tuned Data Detection with a Gaussian Prior}\label{sec:gaus_prior}
We now derive a \mLAMA algorithm variant using a mismatched Gaussian prior when the true signals are taken from a constellation set $\setO$ with equally likely symbols assuming $\Exop_{S_0}[\abs{S_0}^2] = E_s$.
We assume a standard complex Gaussian distribution for the mismatched prior, i.e., $\tilde p(\tilde s_\ell)\sim\setC\setN(0,1)$ as the variance parameter $\tau^t$ will be scaled accordingly to $E_s$ in the tuning stage \fref{eq:tau_opt}.
For the mismatched Gaussian prior, the
message mean function \fref{eq:F} is given by $\mathsf{F}^\text{mm}(r,\tau)= \frac{E_s}{E_s+\tau}r$, \revision{which is a Lipschitz-continuous function}.
\revision{
Substituting $\mathsf{F}^\text{mm}(r,\tau)$ into \fref{eq:SE_MSE} and optimally tuning $\tau^t$, we can derive the following mismatched SE recursion for \fref{thm:SE}:
\begin{align}
\label{eq:gaussPsi}
\sigma_{t+1}^2 &= \No + \beta \notag
\psimmstar( \sigma^2_t )\\
&=
\No + \beta
\min_{\gammasq\geq0}
\!\left\{\frac{E_s^2\sigma_t^2}{(
E_s+\gammasq)^2} + \frac{E_s\gamma^4}{(E_s+\gammasq)^2}\right\}\!.
\end{align}
}
\revision{We note that the SE recursion in \fref{eq:gaussPsi} allows us to compute the decoupled noise variance, i.e., the inverse of post-equalization 
signal-to-interference-and-noise ratio, analytically from the M-LAMA algorithm described in \fref{sec:MCBAMP_alg} without numerical simulations.
By doing so, we can obtain a full performance characterization of M-LAMA's performance in the asymptotic regime, and measure the performance degradation in practical finite-dimensional systems (see \fref{sec:numerical} for the details).
}

The mismatched SE recursion \fref{eq:gaussPsi} only depends on the signal energy $E_s$ and no other properties of the true prior $p(\bms_0)$.
This fact allows us to optimally tune the variance parameters only by knowing $E_s$.
Therefore, if the true prior is unknown, but we know the signal energy, one may use \mLAMA to perform data detection.
Before we proceed to the fixed-point analysis, the following lemma, with proof given in  \fref{app:LAMA_tune_gaussianprior},  connects the tuning stage of \mLAMA in \fref{eq:tau_opt} and  \fref{eq:gaussPsi}.
\begin{lem} 
\label{lem:LAMA_tune_gaussianprior}
Assume a mismatched Gaussian prior $\tilde p(\tilde s)\sim\setC\setN(0,1)$ and the power of the true prior is $\Exop_{S_0}[\abs{S_0}^2] = E_s$. Then, the optimal choice in
the tuning stage \fref{eq:SE_gamma} is $\gammasq_\star^t=\sigma_t^2$ which is the global minimizer to \fref{eq:gaussPsi} for a fixed $\sigma_t^2\geq0$.
\end{lem}
Thus, the mismatched SE recursion \fref{eq:gaussPsi} reduces to
\begin{align}
\sigma_{t+1}^2 = \No + \beta\frac{E_s}{E_s+\sigma_t^2}\sigma_t^2, \label{eq:mmsefixedeq}
\end{align}
and by \fref{lem:unique_fp_MRT}, \mLAMA has a unique fixed point when $\beta\leq1$ regardless of the noise variance $\No$.
Interestingly, the fixed-point equation of \fref{eq:gaussPsi} of this algorithm corresponds to the decoupled noise variance given by the linear MMSE \revision{(L-MMSE)} detector in \cite{TH1999,VS1999}.
If we define signal-to-interference ratio (SIR) as $\text{SIR}=1/\sigma^2$, and let $\tmax\to\infty$, then the fixed-point solution of~\fref{eq:mmsefixedeq} coincides with the SIR given by the linear MMSE detector in the large-system limit \cite{TH1999,VS1999,SV2001}.
Hence, for a mismatched Gaussian prior, \mLAMA achieves exactly the same performance as the linear MMSE detector.
We note that the proofs given in \cite{TH1999,VS1999,SV2001} use results from random matrix theory, whereas our analysis uses the mismatched SE framework proposed in \fref{thm:SE}. Furthermore, our result is constructive, i.e., \mLAMA is a computationally efficient  algorithm that implements linear MMSE detection without the need of computing an explicit matrix inversion.
\subsection{Suboptimal Data Detection with a  Gaussian Prior}\label{sec:ZFMF}
We can replace the optimal tuning stage in \fref{eq:SE_gamma} by a fixed (and predetermined) variance parameter choice for $\gammasq^t$, which leads to a suboptimal, mismatched algorithm, referred to as suboptimal \mLAMA (short \smLAMA).
We now show that this approach leads to other well-known linear data detectors.
In particular, by considering the following two variance parameter choices $\gammasq^t\to0$ and $\gammasq^t\to\infty$ in \fref{eq:SE_gamma}, we obtain the following mismatched SE recursions:
\begin{align*}
(\text{ZF})\,\,\,\sigma_{t+1}^2\! &= \No + \beta\!\!\lim_{\gammasq^t\to0}\!\!\Psi^\text{mm}(\sigma_t^2,\gammasq^t)=\No+\beta\sigma_t^2,\\
(\text{MF})\,\,\,\sigma_{t+1}^2\! &= \No + \beta\!\!\lim_{\gammasq^t\to\infty}\!\!\Psi^\text{mm}(\sigma_t^2,\gammasq^t)
=\No+\beta\Varop_{S_0}[S_0],
\end{align*}
respectively.
As a result of \fref{lem:unique_fp_MRT}, (ZF) and (MF) have a unique fixed point when $\beta<1$ and for any finite $\beta$, respectively, regardless of the noise variance $\No$.
The solution to the fixed-point equation (ZF) when $\beta<1$ and (MF) coincides exactly to the SIR given by ZF and matched filter (MF) detector in the large-system limit  \cite{TH1999,VS1999,SV2001,EC2003}, respectively. Hence, the use of suboptimal variance parameter choices for~$\gammasq^t$ in \smLAMA results in data detectors whose performance matches that of the well-known ZF and MF data detectors.

\section{Mismatched Data Detection with Optimal Tuning: PAM/QAM Constellations} \label{sec:QAMConstellation}

We now propose a variant of the \mLAMA{} algorithm that improves upon \mLAMA{} for Gaussian priors presented in \fref{sec:GaussianPrior} for PAM and QAM constellations, which are frequently used in practice---the premise is to select a mismatched prior that more closely resembles the true prior.
%
Concretely, we will present two algorithms of \mLAMA{} that assume (i) a uniform hypercube prior and (ii) a Gray coding based approximation. For each algorithm variant, we describe the optimal tuning procedure and also present a suboptimal method that avoids parameter tuning.
In what follows, we assume that the true prior is for QAM or PAM constellation sets $\setO$ with equally likely symbols, i.e., $p(s_{0\ell})=\frac{1}{\abs{\setO}} \sum_{a\in\setO}\delta(s_{0\ell}-a)$, where $\abs{\setO}$ is the cardinality of~$\setO$.

\subsection{Optimally-Tuned Uniform Hypercube Prior}\label{sec:softboxlama}
We start by deriving an \mLAMA{} algorithm variant assuming a uniform hypercube prior, which can be visualized by placing a hypercube with length $2\alpha$ around the true prior distribution of square constellations (e.g., QPSK and QAM).
For example, for QPSK, the mismatched hypercube prior corresponds to a uniform distribution on the interval $[-1,+1]$ where $\alpha=1$, rather than using equally-likely symbols from $\{-1,+1\}$, for both the real and imaginary parts.

For the uniform hypercube prior, we use \fref{lem:identical_separable} to compute the posterior mean function independently for the real and imaginary part; the posterior mean function $\mathsf{F}^\text{mm}$ and its first derivative are given by:
\begin{align}\nonumber
\mathsf{F}^\text{mm}(r,\tau) = \, &
\sellr + \frac{\tau}{2} \nu_-(\sellr,\tau/2) \\\label{eq:F_box}
&+
\imageTxt{}\!\left(\selli + \frac{\tau}{2} \nu_-(\selli,\tau/2)
\right)
\\
\nonumber
\mathsf{F'}^\text{mm}(r,\tau) = \, & 1-
\frac{1}{2}
\left(\sellr\nu_-(\sellr,\tau/2) + \alpha \nu_+(\sellr,\tau/2)
\right)\\\nonumber
&-
\frac{1}{2}
\left(\selli\nu_-(\selli,\tau/2) + \alpha \nu_+(\selli,\tau/2)
\right)\\\label{eq:Fprime_box}
&- \frac{\tau}{4}\!\left(
\nu_-^2(\sellr,\tau/2)+\nu_-^2(\selli,\tau/2)
\right),
\end{align}
where we use the following shorthand notations: \mbox{$\sellr=\realpart{r}$}, \mbox{$\selli=\imagpart{r}$}, and
\begin{align*}
\nu_-(r,\tau)=\frac{
e^{-\frac{1}{2\tau}(r+\alpha)^2}-
e^{-\frac{1}{2\tau}(r-\alpha)^2}
}{
\sqrt{2\pi\tau}\!\left(
\Phi\!\left(\frac{r+\alpha}{\sqrt{\tau}}\right)-\Phi\!\left(\frac{r-\alpha}{\sqrt{\tau}}\right)
\right)
},\\
\nu_+(r,\tau)=\frac{
e^{-\frac{1}{2\tau}(r+\alpha)^2}+
e^{-\frac{1}{2\tau}(r-\alpha)^2}
}{
\sqrt{2\pi\tau}\!\left(
\Phi\!\left(\frac{r+\alpha}{\sqrt{\tau}}\right)-\Phi\!\left(\frac{r-\alpha}{\sqrt{\tau}}\right)
\right)
}.
\end{align*}
\revision{It can be shown that $\mathsf{F'}^\text{mm}$ is bounded above to establish its Lipschitz continuity.}
The mismatched SE recursion is obtained by \fref{thm:SE} and can be evaluated numerically.

This mismatched data \revision{detection} algorithm suffers from two main disadvantages in practical systems:%
\begin{inparaenum}[(i)]
\item The \mLAMA{} algorithm with a hypercube prior is not efficient from a hardware perspective as the function in \fref{eq:F_box} involves transcendental functions.
In fact, this algorithm must evaluate the functions $\nu_+(r,\tau)$ and $\nu_-(r,\tau)$ in every iteration and for every antenna, which require---similar to that of the optimal \LAMA algorithm~\cite{JGMS2015conf}---high numerical precision and a large dynamic range; see, e.g.,~\cite{BBBBQS2014} for a detailed discussion of implementation aspects.
\item The tuning stage in \fref{eq:tau_opt} turns out to be non-trivial---while a grid search or bisection method are viable  methods to find a minimum numerically, implementing such methods in hardware is impractical.
\end{inparaenum}

\subsection{Suboptimally-Tuned Uniform Hypercube Prior}\label{sec:sblama}

Analogously to the ZF detector in \fref{sec:ZFMF}, which used a suboptimal tuning parameter, we can derive a sub-optimal variant of \mLAMA (\smLAMA) with the uniform hypercube prior from \fref{sec:softboxlama}, where we replace the tuning stage in \fref{eq:tau_opt} by the fixed choice $\tau^t\to0$.
This suboptimal, but fixed, choice leads to a much simpler algorithm compared to the optimally-tuned \mLAMA algorithm and in addition, makes the performance analysis more accessible.
First, the posterior mean function simplifies to
\begin{align}
& \lim\limits_{\tau\to0} \mathsf{F}^\text{mm}(r,\tau) = \sellr+ \sign(\sellr)\min\!\left\{\alpha-\abs{\sellr},0\right\} \notag \\
& \qquad \qquad \qquad +\imageTxt{}
\left(
\selli+ \sign(\selli)\min\!\left\{\alpha-\abs{\selli},0\right\}\right)\!, \label{eq:F_SMLAMA}
\end{align}
which can be evaluated efficiently. \revision{
Compared to \fref{eq:F_box} and~\fref{eq:Fprime_box}, computing \fref{eq:F_SMLAMA} is much simpler as it does not require the computation of (i) a sum and difference of Gaussian PDFs (which requires exponential functions) and (ii) the inverse of difference of Gaussian CDFs. Instead, minimum and sign operations can be computed efficiently in hardware as they merely require a subtraction of two numbers.
}
Second, computing $\lim_{\tau\to0}\mathsf{F'}^\text{mm}(r,\tau)$ \revision{is straightforward, which is simply given by the following result:}
\begin{align*}
\lim_{\tau\to0}\mathsf{F'}^\text{mm}(r,\tau) =\textstyle \frac{1}{2}\mathbb{I}(\abs{\sellr}<\alpha)+\frac{1}{2}\mathbb{I}(\abs{\selli}<\alpha).
\end{align*}
Third, by letting $\tau\to0$, the tuning stages in \fref{eq:decouple_estimate} and \fref{eq:tau_opt} are no longer required as the variance parameter $\tau$ is fixed.

Since $\mathsf{F}^\text{mm}$ is composed of piece-wise linear functions, we can explicitly state the mismatched SE recursion in \fref{eq:SE_MSE} for \smLAMA
under $M^2$-QAM constellations with the aid of \fref{lem:identical_separable} (see \fref{app:qam_SE_deriv} for the derivation):
\begin{align}\nonumber
\sigma_{t+1}^2 =\, &
\No + \beta\Psi^\text{mm}(\sigma_t^2)\\\nonumber
=\, &
\No + \beta
\frac{2}{M}\sum_{k=1}^{M/2}
\Bigg[
\sigma^2_t +
\left(2 \bar\alpha_k^2-\sigma_t^2
\right)
Q\!\left(
\frac{\bar\alpha_k}{\sigma_t/\sqrt{2}}
\right)
\\\nonumber
&-\frac{\sigma_t}{\sqrt{\pi}}
\!\left(\bar\alpha_k
\exp\!\left(
-\frac{\bar\alpha_k^2}{\sigma_t^2}\right)
+\alpha_k\exp\!\left(
-\frac{\alpha_k^2}{\sigma_t^2}\right)
\right)\!\\\label{eq:qam_SE_deriv}
&+
\!\left(2\alpha_k^2 - \sigma_t^2
\right)
Q\!\left(\frac{\alpha_k}{\sigma_t/\sqrt{2}}\right)\!
\Bigg],
\end{align}
where $\alpha=M-1$, $\bar\alpha_k = \alpha-(2k-1)$ and $\alpha_k =\alpha+(2k-1)$.

We now present conditions on the system ratio $\beta$ where \smLAMA has a unique fixed point. The following \fref{lem:ERT_QAM}, with proof in \fref{app:ERT_QAM}, shows that the MRT of \smLAMA for $M^2$-QAM is given by $\betamin=(1-1/M)^{-1}$.
\begin{lem}\label{lem:ERT_QAM} Assume that $S_0$ is selected from $M^2$-QAM with equally likely symbols. Then, the minimum recovery threshold (MRT) \cite{JGMS2015conf} for
\smLAMA is given by $\betamin=(1-1/M)^{-1}$.
Moreover, \fref{lem:unique_fp_MRT} holds for \smLAMA at $\beta=\betamin$, i.e., \smLAMA has unique fixed point at $\beta=\betamin$ regardless of the noise variance $\No$.
\end{lem}

With \fref{lem:ERT_QAM}, we also obtain the same MRT for $M$-PAM by \fref{lem:identical_separable}. We omit the proof and refer to \cite{JGMS2015}.

\begin{cor}\smLAMA has the same MRT for $M^2$-QAM and $M$-PAM in a real-valued system.
\end{cor}

We now show that this \smLAMA variant achieves the same performance as a well-known relaxation of the maximum likelihood data detection problem \cite{tan2001constrained,yener2002cdma,pan2014mimo,TAXH2015}.
This algorithm, known as box-relaxation (BOX, for short) detector, solves the following convex problem \cite{SGYB14,TAXH2015,TAH2018,SJS2017,ASH2019}:
\begin{align}
 \hat\vecs = \argmin_{\tilde\bms\in\complexset^{}} \, \|\bmy-\bH\tilde\bms\|_2 \quad \text{subject\,\,to}\,\, \|\tilde\bms\|_\infty\leq \alpha \label{eq:boxproblem}
\end{align}
and slices the individual entries of $\hat\vecs$ onto the $M^2$-QAM (or $M$-PAM) constellations. The next result shows that  \smLAMA achieves the same error-rate performance as the BOX detector, while providing a simple and computationally efficient algorithm. The proof is given in \fref{app:hassibi}.
\begin{lem} For a complex-valued MIMO system with $M^2$-QAM constellations (or a real-valued MIMO system with $M$-PAM constellations), \smLAMA achieves the same performance as the BOX detector  \cite{TXH2018,ASH2019} in the large-system limit.\label{lem:hassibi}
\end{lem}

We emphasize that in precise performance analysis of the BOX detector for real-valued systems was shown in \cite{TXH2018}, and was extended to that of complex-valued systems in \cite{ASH2019}, whereas our analysis connects both complex-valued $M^2$-QAM and real-valued $M$-PAM systems via SE.
In addition, the authors in \cite{ASH2019} were able to recover identical results of the MRT of $M^2$-QAM shown in \fref{lem:ERT_QAM}.
Moreover, we note that our analysis is constructive, i.e., the \smLAMA{} algorithm can be used to detect both QAM and PAM constellations.
\subsection{Optimally-Tuned Gray-Coding-Based Approximation}

We note that for certain systems, it may be of interest to directly postulate a posterior mean function, rather than  assuming a mismatched prior distribution in the first place.
In this section, we will derive an \mLAMA algorithm variant that exploits Gray mapping (from bits to constellation points), which is used in many practical communication systems.
For the sake of brevity, we will derive the posterior mean function and analysis for 16-QAM and uniform priors only. However, the proposed approach can easily be generalized to higher-order QAM constellations and non-uniform priors.
We first start by noting that the posterior mean function \fref{eq:FrealFunction} for 16-QAM for uniform priors corresponds to
\begin{align*}
\mathsf{F}^\textnormal{16-QAM}(r,\tau) = \mathsf{F}^\textnormal{4-PAM}(\sellr,\tau/2) + \imageTxt{}\mathsf{F}^\textnormal{4-PAM}(\selli,\tau/2),
\end{align*}
where we used the separability property of 16-QAM in \fref{def:sep_const}.
We now introduce the following shorthand notation for the posterior mean for 4-PAM:
\begin{align}
\mathsf{F}^\textnormal{4-PAM}(r,\tau)\label{eq:F_originalLAMA}
&=
\frac{
-3e_{-3}
-1
e_{-1}
+  e_{1}
+3e_3
}
{
e_{-3}
+e_{-1}
+  e_{1}
+e_3
},
\\
\label{eq:e_a}
e_a &= \exp \!\left( -\frac{1}{2\tau}(r-a)^2)\right)\!.
\end{align}
Here, we omit $r$ and $\tau$ from $e_a(r,\tau)$ for the ease of notation.
We note that the function $\mathsf{F}^\textnormal{4-PAM}$ can be rewritten as
\begin{align*}
\mathsf{F}^\textnormal{4-PAM}(r,\tau)  = \sum_{a\in\setO^\textnormal{4-PAM}} w_a(r,\tau) a,
\end{align*}
where $\setO^\textnormal{4-PAM}= \{\pm3,\pm1\}$ and $w_a(r,\tau)$ is a weight distribution so that $\sum_{a\in\setO^\textnormal{4-PAM}} w_a(r,\tau) = 1$.
For 4-PAM, we have
\begin{align}
\label{eq:weight_comp}
w_a(r,\tau) =
\frac{
e_a}
{
e_{-3}
+e_{-1}
+  e_{1}
+e_3 }.
\end{align}
The high arithmetic precision requirement of computing the posterior mean mainly stems from the computation of \fref{eq:weight_comp};
this is due to the fact that $e_a$ decays exponentially fast to zero for small $\tau$. Thus, computing \fref{eq:weight_comp} requires excessively high numerical precision, which makes the design of efficient integrated hardware implementations challenging.

We now propose an approximation of \fref{eq:weight_comp} that not only alleviates the arithmetic precision requirements, but also achieves better performance than the hypercube prior discussed in \fref{sec:softboxlama}.
To do so, we exploit Gray coding \cite{paulraj03}, which is used in most communication standards.

We begin by assuming that all the bits mapped to a constellation point are independent. Given this assumption, we can decompose the symbol-domain weight distribution in~\fref{eq:weight_comp} into products of bit-domain probabilities as
\begin{align*}
w_{-3} &= (1-p_1)(1-p_0),\quad
 w_{-1}  = (1-p_1)p_0,\quad
\\
w_{1} &= p_1p_0,\qquad\qquad\qquad\,\,\,\,
 w_{3}  = p_1(1-p_0),
\end{align*}
where we omit the indices $(r,\tau)$ for simplicity.
Here, we introduced $p_b$ for $b\in\{0,1\}$ that represent the probability that $b$th bit is equal to 1.
For Gray coding and bit-wise independence assumption, we have $p_1=w_1 + w_3$ and $p_0= w_{-1} + w_1$.
Thus, we can simplify $\mathsf{F}^\textnormal{4-PAM}_\textnormal{Gray}(r,\tau) = (2p_0 -3)(1-2p_1)$.

Now, instead of computing $p_1$ and $p_0$ directly, we compute the log-likelihood ratio (LLR) $\Lambda_b = \log\big(\frac{p_b}{1-p_b}\big)$ and use the relation $p_{b} = \frac{1}{2}(1 + \tanh(\frac{1}{2}\Lambda_{b}))$ for bits $b=0,1$, which can be computed efficiently and in a numerically stable manner in hardware via look-up tables \cite{studer2011asic}.
With this formulation, the derivative becomes
\begin{align*}
\mathsf{F'}^\textnormal{4-PAM}_\textnormal{Gray}(r,\tau)
=\,&
(2-\tanh(\bar\Lambda_0))
(1-\tanh^2(\bar\Lambda_1))
\bar\Lambda_1' \notag
\\
&-
(1-\tanh^2(\bar\Lambda_0))\bar\Lambda_0'
\tanh(\bar\Lambda_1)
\\
=\, &  4(3-2p_0)
p_1(1-p_1)\bar\Lambda_1' \notag \\
& +
4(1-p_0)(1-2p_1)p_0\bar\Lambda_0'
,
\end{align*}
where we use the shorthand notation $\bar\Lambda_b = \Lambda_b/2$ for both $b=0,1$, and $\bar\Lambda_b'$ is the first derivative of $\bar\Lambda_b$ \revision{with respect to}~$r$.
Given $\bar\Lambda_b'$ and the bit probabilities $p_b$, the above expression of $\mathsf{F}'$ can be evaluated in a straightforward manner.
\revision{Since $|\tanh(\cdot)|\leq1$, we can see that $\mathsf{F'}^\textnormal{4-PAM}_\textnormal{Gray}(r,\tau)$ is bounded and thus, Lipschitz continuous.}

\subsubsection{Computation of the Log-Likelihood Ratio}
We now  elaborate how to compute the log-likelihood ratio (LLR) values~$\Lambda_0$ and~$\Lambda_1$.
Again, we assume 16-QAM.
We propose two methods to compute $\Lambda_0$ and $\Lambda_1$: (i) a direct approach and (ii) a low-complexity approach via the max-log approximation.

(i) A direct computation of LLR values is straightforward and is computed by noting the fact that $p_1 = w_1+w_3$ and $p_0 = w_{-1} + w_1$.
Based on $p_0$ and $p_1$, we can rewrite the LLR expressions by using $w_a$ in \fref{eq:weight_comp} and  $e_a$ defined in \fref{eq:e_a} as
\begin{align*}
\Lambda_0 &= \log\!\left( \frac{p_0}{1-p_0}\right) = \log \!\left(\frac{w_{-1}+ w_1}{ w_{-3} + w_{3}}\right)
\\
&= \log \!\left(\frac{e_{-1}+ e_1}{ e_{-3} + e_{3}}\right)
=8\rho + \log \!\left(
 \frac{\cosh(2\rho r)}{\cosh(6\rho r)}
\right)\!,
\\
\Lambda_1 &= 8\rho r + \log
\!\left(
\frac{\cosh(2\rho(r-2))}{\cosh(2\rho(r+2))}
\right)\!,
\end{align*}
where we defined the shorthand notation $\rho=1/\tau$.

(ii) Although the expressions for $\Lambda_0$ and $\Lambda_1$ are straightforward, computing them in hardware can be  challenging due to the transcendental nature of the ratio term of hyperbolic cosine functions.
Thus, to simplify hardware designs, we propose a low-complexity method to approximate the exact LLRs $\Lambda_0$ and~$\Lambda_1$ via the max-log approximation \cite{studer2011asic}.
The idea is to realize that $\log(\exp(a) + \exp(-a)) = a + \log( 1 + \exp(-2\abs{a}))\simeq a $ for large values of $a$ as $\exp(-2\abs{a})$ quickly converges to 0.
Thus, applying the max-log approximation to $\Lambda_0$ results in
\begin{align}
\log
\!\left(
\frac{\cosh(2\rho r)}{\cosh(6\rho r)}
\right)
&= \log
({\cosh(2\rho r)}) - \log({\cosh(6\rho r)})
\\
&= 2 \rho \abs{r} + \log(1 + \exp(-4\rho\abs{r})) \notag
\\&\quad
-  6 \rho \abs{r} - \log(1 + \exp(-12\rho\abs{r}))
\\
& \stackrel{(a)}{\simeq} 2 \rho \abs{r} - 6\rho\abs{r} = -4\rho\abs{r}\!.
\end{align}
Similarly, for $\Lambda_1$ we have
\begin{align}
\log \!\left(\frac{\cosh(2\rho(r-2))}{\cosh(2\rho(r+2))}\right)
&\stackrel{(b)}{\simeq} 2\rho\abs{r-2} - 2\rho\abs{r+2}\!.
\end{align}
We note that in the approximations $(a)$ and $(b)$, we have ignored the correction term $\log(1+\exp(-c\abs{r}))$; one may include an approximated value of the correction term to mitigate the loss of max-log approximation at a moderate overhead in complexity.
However, we will show in \fref{sec:numerical} that the proposed max-log approximation does not result in an error-rate  performance  loss.
The  max-log LLR values $\Lambda_0$ and $\Lambda_1$ are given by
\begin{align*}
\Lambda_0^\text{max-log}(r,\rho^{-1}) &= 4\rho(2-\abs{r}),
\\
\Lambda_1^\text{max-log}(r,\rho^{-1})
&= 2\rho( 4r + \abs{r-2} - \abs{r+2})
\\&=
\begin{cases}
8\rho(r+1), & r\in(-\infty,-2), \\
4\rho r, & r\in [-2,2], \\
8\rho (r-1), & r\in(2,\infty). \\
\end{cases}
\end{align*}
\revision{
We note that computing the resulting LLR values does not require any transcendental functions and only requires boundary checks and multiplications, which can be implemented efficiently in hardware. }

\subsection{Suboptimal Tuning of the Gray-coding Based Approximation}

As described in \fref{sec:optimal_tuning}, optimal tuning minimizes the performance loss of mismatched priors.
Optimal tuning for the Gray-coding based approximation requires one to solve the optimization problem in \fref{eq:SE_gamma}.
For the Gray-coding based approximation, the posterior mean with the exact LLR computation is given by
\begin{align}\notag
&\mathsf{F}^\text{4-PAM}_\text{Gray}(r,\rho^{-1} ) =
( 2-\tanh(\bar\Lambda_0))\tanh(\bar\Lambda_1)
\\
&\quad\qquad =
\left(2- \tanh\!\left(
4\rho + \frac{1}{2}
\log \frac{\cosh(2\rho r)}{\cosh( 6\rho  r)}
\right)\!
\right)\! \notag
\\
&\quad\qquad\quad\times
\tanh\!\left(4\rho r + \frac{1}{2}\log \frac{\cosh(2\rho (r-2))}{\cosh( 2\rho (r+2))}\right)\!.
\label{eq:F_16qam_no_maxlog}
\end{align}
\revision{We note that the Gray-coding based approximation \fref{eq:F_16qam_no_maxlog} substantially mitigates numerical precision requirements compared to the exact posterior mean from \fref{eq:F_originalLAMA}, as no exponential functions and the inverse of their sum are required.
However, evaluating~\fref{eq:F_16qam_no_maxlog} still requires the computation of ratios of $\cosh(\cdot)$. These calculations can be avoided by using the max-log approximation \cite{studer2011asic}
 as follows:
}
\begin{align}\label{eq:F_16qam_maxlog}
\mathsf{F}^\text{4-PAM}_\text{max-log}(r,\rho^{-1} ) &=
\left(
2-
\tanh\!\left(
2\rho (2 - \abs{r})
\right)\!
\right)\! \notag
\\
&\quad\times
\tanh\!\left(
\rho( 4 r + \abs{r-2} - \abs{r+2})
\right)\!.
\end{align}
\revision{
Compared to  \fref{eq:F_originalLAMA} and \fref{eq:F_16qam_no_maxlog}, the max-log approximation  in \fref{eq:F_16qam_maxlog} significantly reduces hardware complexity as no $\cosh(\cdot)$ and their ratios are required, and $\cosh(\cdot)$ is approximated by sums of differences of absolute values.
We will show in \fref{sec:numerical} that this proposed max-log approximation yields virtually no loss in error-rate performance while reducing complexity.
}

\revision{We note that solving the optimization problem in \fref{eq:SE_gamma} with the posterior mean functions above is difficult.}
Analogously to optimal tuning for the uniform hypercube prior in \fref{sec:softboxlama}, a grid search or bisection methods would be viable methods but are impractical in hardware.
\revision{
Based on numerical calculations for optimization of \fref{eq:SE_gamma} for 16-QAM constellation, and $M^2$-QAM constellations, more generally, we have observed that using a suboptimal choice of the tuning parameter and simply setting \mbox{$\tau =\rho^{-1} =  \sigma^2$} results in excellent performance that performs extremely close to the optimally-tuned value $\tau$ without any overhead in complexity.
}


\begin{figure*}[tp]
 \centering
\subfigure[$\beta=0.5$ with QPSK]{
\includegraphics[width=0.4\textwidth]{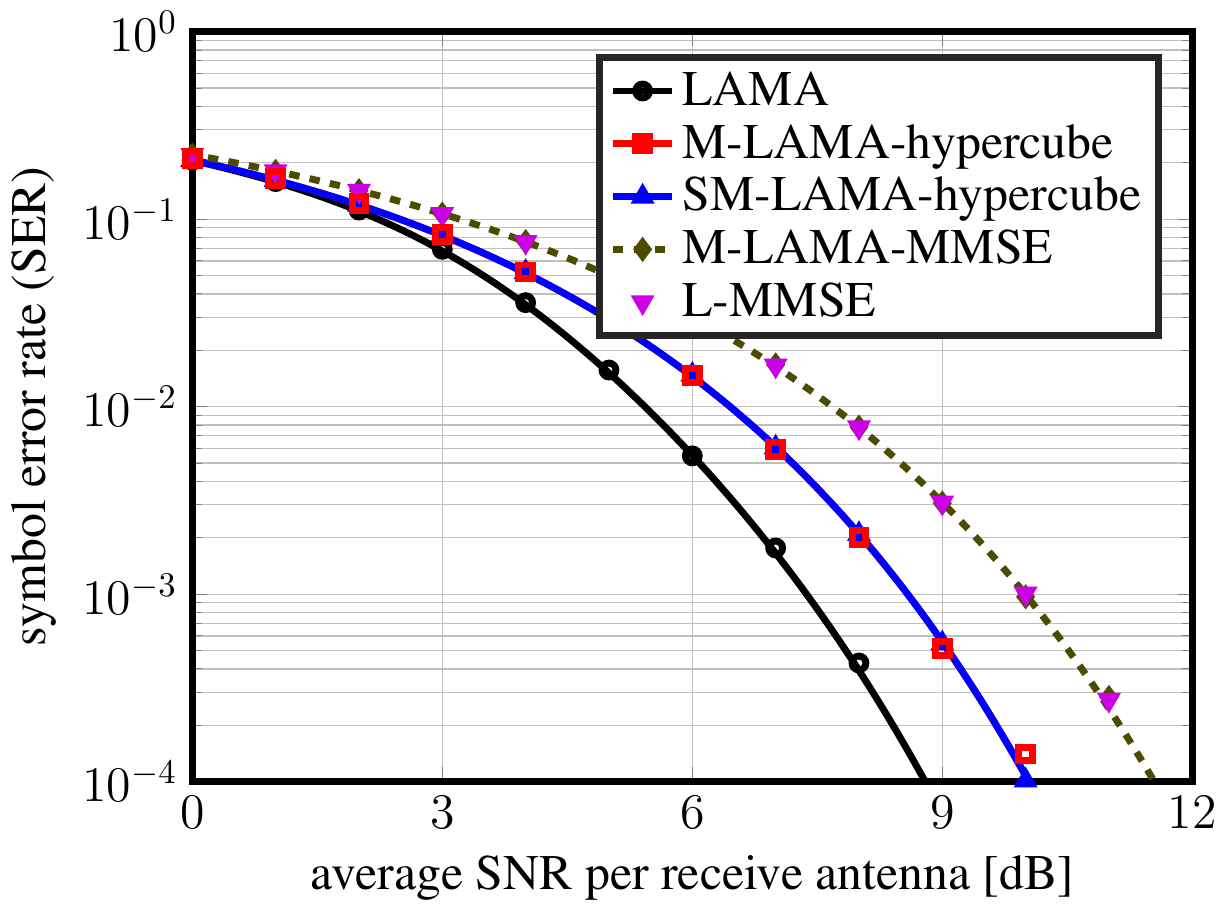}
\label{fig:SE_QPSK}
}
\hspace{1cm}
\hspace{0.1cm}
\subfigure[$\beta=0.5$ with 16-QAM]{
\includegraphics[width=0.4\textwidth]{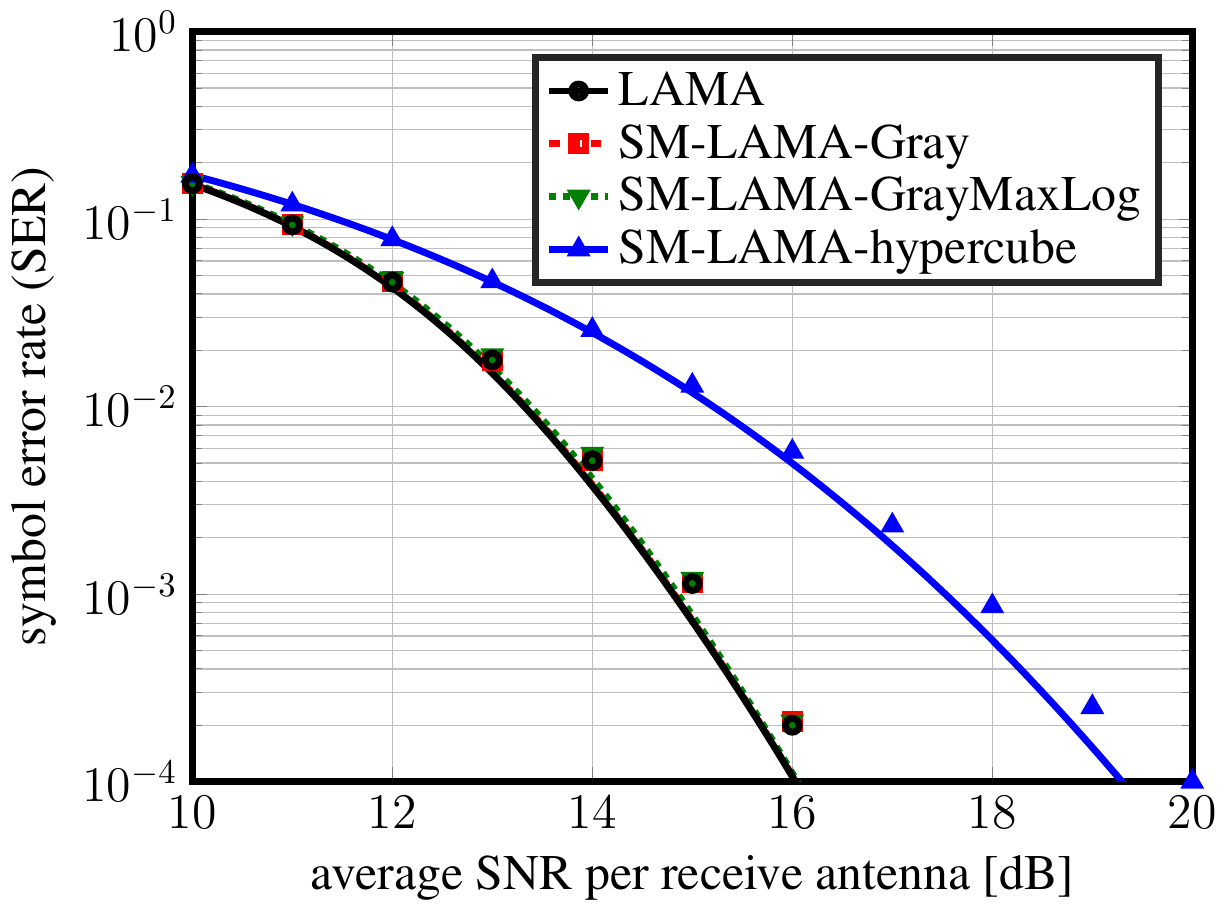}
\label{fig:SE_QAM16}
}
\caption{
Symbol-error rate plots for \mLAMA algorithm and its variants for
a $128\times64$ massive MU-MIMO system with 10 algorithm iterations.
The error-rate performance predicted by the SE framework in the large-system limit are
shown with lines whereas the error-rate performance obtained by numerical simulations are
shown with markers.
For QPSK, \smLAMA with the uniform hypercube prior performs within 1\,dB from LAMA \cite{JGMS2015conf}, which achieves IO performance in the large-system limit.
For 16-QAM, \smLAMA with the Gray-coding based approximation performs on part with LAMA.
Thus, \mLAMA achieves near-IO error-rate performance at a lower computational complexity by using a carefully designed mismatched prior.
}
\label{fig:error_rate}
\vspace{-0.2cm}
\end{figure*}

\section{Numerical Results}\label{sec:numerical}

We now compare the error-rate performance of the proposed \mLAMA algorithm variants in \fref{fig:error_rate} \revision{via Monte--Carlo simulations}.
Although the mismatched SE framework  \fref{thm:SE} enables an exact error-rate analysis in the large-system-limit, we also provide numerical simulations in a finite dimensional massive MU-MIMO system with $\MR=128$ basestation and $\MT=64$ single-antenna users  for two constellations: QPSK and 16-QAM.
The error-rate performance predicted by the SE framework in the large-system limit is shown with either solid, dashed, or dotted lines, whereas the error-rate performance obtained by numerical simulations are represented by the  markers.
For both simulations, we also compare our results to the \LAMA{} algorithm proposed in \cite{JGMS2015conf}, which was shown to achieve near-IO performance in finite systems and IO performance in the large-system limit for $\beta=0.5$ and for both constellations.

\fref{fig:SE_QPSK} shows the error-rate performance of \mLAMA with QPSK.
We show the performance of three mismatched LAMA algorithms: optimally-tuned \mLAMA with a Gaussian prior (called ``\mLAMA-MMSE'') as well as optimally- and sub-optimal tuned \mLAMA for hypercube prior; we also include the error-rate performance of \LAMA as a baseline.
We exclude the Gray-coding based approximation as it is optimal for QPSK. In other words, for QPSK, the Gray-coding based approximation achieves the same error-rate performance as \LAMA.
We see that the proposed \mLAMA algorithms achieve similar error-rate performance in the finite-dimensional system as the predicted error-rate performance in the large-system limit.
In particular, \mLAMA with an optimally-tuned Gaussian prior (\mLAMA-MMSE) achieves near-identical error-rate performance to the exact linear MMSE detector, which agrees with our SE analysis.
The optimally-tuned \mLAMA{} algorithm and its ZF variant \smLAMA{} for the uniform hypercube prior performs within 1\,dB of LAMA \cite{JGMS2015conf}.
Furthermore, as noted in \fref{lem:hassibi}, SM-LAMA-hypercube achieves identical error-rate performance as that given by the BOX detector, whose performance was analyzed before in \cite{SJS2017,TXH2018,ASH2019,HL2020}.

\fref{fig:SE_QAM16} shows the error-rate performance of \mLAMA with 16-QAM.
We show the performance of both Gray-coding based approximations as well as the sub-optimally tuned \mLAMA algorithm with a hypercube prior.
Compared to \fref{fig:SE_QPSK}, we observe a slight performance gap between the asymptotic predictions by SE and numerical simulations---note that the gap disappears when increasing the system dimension.
Among the simulated methods, we first observe that the Gray-coding based approximation provides significant performance gains compared to the hypercube-prior.
Our results also show that there exist no performance loss between the Gray-coding-based and the max-log approximation, compared to the original \LAMA{} algorithm that requires repeated evaluations of \fref{eq:FrealFunction}.
Clearly, \fref{fig:SE_QAM16} demonstrates that \mLAMA{} with carefully designed mismatched priors is able to achieve near-optimal error-rate performance at (often significantly) lower complexity and without the need for complicated transcendental functions that prevent efficient hardware designs---our recent ASIC prototype \cite{JCS2019}, which uses the max-log Gray coded mismatched function, demonstrates the real-world efficacy of  \mLAMA{}.

\section{Conclusions}

We have presented the \mLAMA{} algorithm along with the mismatched SE recursion.
We have shown that for a mismatched Gaussian prior, optimally-tuned \mLAMA{} and suboptimally tuned {\smLAMA} achieve the same performance as the linear MMSE,  ZF, and MF detectors.
For a QAM constellations, we have presented two variants of \mLAMA and characterized the performance for a uniform hypercube prior and a Gray-coding based approximation.
For a mismatched uniform hypercube prior, we have established conditions on the system ratio $\beta$ for which \smLAMA{} has a unique fixed point under $M^2$-QAM constellation. In addition, we have shown that the \mLAMA{} algorithm achieves identical error-rate performance compared to convex-relaxation methods.
Although the presented theoretical results are only valid in the large-system limit, our simulations have shown that \mLAMA and its variants achieve near-IO performance in realistic, finite dimensional massive MIMO systems.

There are multiple avenues for future work.
Our analysis pertains to the large-system limit---a theoretical study in the finite dimensional setting, e.g., using tools from \cite{RV2016}, would allow a more accurate performance prediction in finite-dimensional massive MU-MIMO systems.
A precise analysis of the incorrectly-decoded number of bits for mismatched detectors, as developed recently in~\cite{HL2020} for the box-relaxation detector, is left for future work.
Furthermore, exploring the fundamental connection between {\smLAMA} and certain convex optimization formulations, such as the box-relaxation in \fref{eq:boxproblem}, is interesting in its own right.

\section*{Acknowledgments}
The authors would like to thank Ramina Ghods for discussions on AMP and state evolution with Gaussian priors.
\revision{The work of~C. Studer was supported in part by ComSenTer, one of six centers in JUMP, a Semiconductor Research Corporation (SRC) program sponsored by DARPA, by an ETH Research Grant, and by the US NSF under grants CNS-1717559 and ECCS-1824379.}

\appendices

\section{Proof of \fref{lem:lama_already_opt}}\label{app:lama_already_opt}

Since there is no prior mismatch, the conditional mean $\sigma_t^2=\gammasq^t$ minimizes the MSE, where the MSE is equivalent to the conditional variance \cite{GWSS2011}. As a result, we have
\begin{align*}
\min_{\gammasq\geq0}\Psi^\text{mm}(\sigma_t^2,\gammasq)
=\Exop_{S_0,Z}\!\left[\abs{\mathsf{F}(S_0+\sigma_tZ,\sigma_t^2)-S_0}^2\right]\!.
\end{align*}
Therefore, \fref{eq:SE_gamma} and \fref{eq:SE_MSE} are equivalent and reduces to the SE recursion given by IO-LAMA in \cite{JGMS2015conf}.

\section{Proof of \fref{lem:LAMA_tune_gaussianprior}}\label{app:LAMA_tune_gaussianprior}

The proof is similar to the steps in \cite{MMB2015} to show that $\Psi^\text{mm}(\sigma_t^2,\gammasq^t)$ is quasi-convex in $\gammasq^t$.
To show the quasi-convexity, we will show that $\frac{\dd}{\dd\gammasq^t}\Psi^\text{mm}(\sigma_t^2,\gammasq^t)$ has only one sign-change.
The proof is straightforward as
\begin{align*}
\frac{\dd}{\dd\gammasq^t}\Psi^\text{mm}(\sigma_t^2,\gammasq^t) = 2\!\left(\frac{E_s}{E_s+\gammasq^t}\right)^{\!3}(\gammasq^t - \sigma_t^2),
\end{align*}
so $\frac{\dd}{\dd\gammasq^t}\Psi^\text{mm}(\sigma_t^2,\gammasq^t)$ has one sign-change at $\gammasq^t=\sigma_t^2$. Note that $\frac{\dd}{\dd\gammasq^t}\Psi^\text{mm}(\sigma_t^2,\gammasq^t)\big\vert_{\gammasq^t\rightarrow0}<0$ so $\tau^t =\sigma_t^2$ is the global minimizer for $\Psi^\text{mm}(\sigma_t^2,\gammasq^t)$.

\revisionthree{
\section{Proof of \fref{lem:lem_converge}}\label{app:proof_lemma_convergence}}
\revisionthree{
In order to prove \fref{lem:lem_converge}, we borrow results from \cite[Thm.~2]{BM2011} where we assume that $\mathsf{F}^\text{mm}$ is a Lipschitz-continuous function for the mismatched prior.
The proof is straightforward by first setting $\psi(a,b)=|a-b|^2$ in~\cite[Eq. 3.7]{BM2011} and noting that $b_i^t-w_i$ in~\cite[Eq.~3.10]{BM2011} corresponds to the residual~$\bmr^t$ in step~\fref{eq:Req}.
For the LHS of \cite[Eq. 3.7]{BM2011}, we have
\begin{align*}
\lim_{\MR\to\infty}\frac{1}{\MR} 
\sum_{k=1}^\MR \psi(r_k^t - n_k,n_k) 
= 
\lim_{\MR\to\infty}\frac{1}{\MR} 
\sum_{k=1}^\MR |r_k^t|^2.
\end{align*}
For the RHS, we first define  $Z\sim\setC\setN(0,1)$ and (scalar) noise $\bar{N}\sim\setC\setN(0,\No)$, and use $\gamma_t^2 =\beta\Psi^\textnormal{mm}(\sigma_{t-1}^2,\tau^{t-1})$ so that
\begin{align*}
\Exop_{Z,\bar{N}}\!\left[ \psi( \gamma_t Z, \bar{N} ) \right]& =
\Exop_{Z,\bar{N}}\!\left[ | \gamma_t Z - \bar{N} |^2 \right]
\\
&= \No + \beta\Psi^\textnormal{mm}(\sigma_{t-1}^2,\tau^{t-1} ) = \sigma_{t}^2,
\end{align*}
where the last equality is an immediate results of~\fref{eq:SE_MSE}. 
}

\section{Derivation of \fref{eq:qam_SE_deriv}}\label{app:qam_SE_deriv}

We will compute \fref{eq:qam_SE_deriv} by first evaluating the mismatched SE recursion for a $M$-PAM system (under real-valued noise) with equally likely priors and then, use \fref{lem:identical_separable} to express the relation for $M^2$-QAM.
We start with the sub-optimally tuned posterior mean \fref{eq:F_SMLAMA}, which we define as
\begin{align*}
\mathsf{F}^\alpha( r )=
\begin{cases}
-\alpha, & r \in (-\infty,-\alpha),\\
r, & r\in[-\alpha,\alpha],\\
+\alpha, & r \in (\alpha,\infty).
\end{cases}
\end{align*}
Note that for equally likely priors, the $M$-PAM constellation can be expressed by $p(s_\ell)=\frac{1}{M}\sum_{k=-M/2+1}^{M/2}\delta(s_\ell-(2k-1))$.
Then, for a given $\sigma$ and $Z\sim\setN(0,1)$, for some symbol $2k-1$
we have the following expression:
\begin{align*}
&\Exop_{Z}[(\mathsf{F}^\alpha((2k-1)+\sigma Z)-(2k-1))^2]\\
&\quad =\sigma^2 + (\bar\alpha_k ^2-\sigma^2)
Q\!\left(
\frac{\bar\alpha_k}{\sigma}
\right)+ (\alpha_k^2-\sigma^2)
Q\!\left(
\frac{\alpha_k }{\sigma}
\right) \notag
\\&\quad -
\frac{\sigma}{\sqrt{2\pi}} \bar\alpha_k
\exp\!\left(
-\frac{\bar\alpha_k^2}{2\sigma^2}\right)
-
\frac{\sigma}{\sqrt{2\pi}}\alpha_k
\exp\!\left(
-\frac{\alpha_k ^2}{2\sigma^2}\right)\!,
\end{align*}
where we denote $\bar\alpha_k = \alpha-(2k-1)$, and $\alpha_k = \alpha+(2k-1)$.
Thus, by exploiting symmetry of $M$-PAM, we have
\begin{align}\nonumber
&\Psi^\text{PAM}(\sigma^2) =
\frac{2}{M}\sum_{k=1}^{M/2}
\Exop_{Z}[(\mathsf{F}^\alpha((2k-1)+\sigma Z)-(2k-1))^2] \\\nonumber
&=
\sigma^2 + \frac{2}{M}\sum_{k=1}^{M/2}
\Bigg[
(\bar\alpha_k ^2-\sigma^2)
Q\!\left(
\frac{\bar\alpha_k}{\sigma}
\right)\! + (\alpha_k^2-\sigma^2)
Q\!\left(
\frac{\alpha_k }{\sigma}
\right)\!
\\&-\label{eq:PAM_PSI}
\frac{\sigma}{\sqrt{2\pi}} \bar\alpha_k
\exp\!\left(
-\frac{\bar\alpha_k^2}{2\sigma^2}\right)
-
\frac{\sigma}{\sqrt{2\pi}}\alpha_k
\exp\!\left(
-\frac{\alpha_k ^2}{2\sigma^2}\right)\!\Bigg].
\end{align}
From $\Psi^\text{PAM}(\sigma^2)$, we can obtain  $\Psi^\text{QAM}$  via \fref{lem:identical_separable} by rewriting
 $\Psi^\text{QAM}(\sigma^2)=2\Psi^\text{PAM}(\sigma^2/2)$ with $\alpha=M-1$.

\section{Proof of \fref{lem:hassibi}}\label{app:hassibi}

We start with the following result from \cite{TXH2018} that establishes the error-rate performance of the BOX detector.

\begin{thm}[Thm 3.1 \cite{TXH2018}]\label{thm:hassibi}
Assume a real-valued $M$-PAM system with $\beta<(1-1/M)^{-1}$.
The symbol-error rate in the large-system limit converges to $2(1-1/M)Q(1/\sigma_\star)$, where $\sigma_\star$ is the unique minimizer to $F_M(\sigma)$:
\begin{align}
\label{eq:hassibi_FP}
F_M(\sigma)&=
\frac{\sigma}{2}\!\left(\frac{1}{\beta}-\frac{M-1}{M}\right)\!
+
\frac{\No}{2\beta\sigma}
+
\frac{1}{M}
\sum_{k\in K}
S(\sigma,k),
\end{align}
where $K = \{2,4,\ldots,2(M-1)\}$, and
\begin{align*}
S(\sigma,k) = \! \left(\sigma +  \frac{k^2}{\sigma}\right)\! Q\!\left(\frac{k}{\sigma}\right)\! - \frac{k}{\sqrt{2\pi}} \exp\!\left( - \frac{k^2}{2\sigma^2}\right)\!.
\end{align*}
\end{thm}
Compared to the exact expression in  \cite{TXH2018}, we have an additional $\beta$ term in the denominator of $\frac{\No}{2\beta\sigma}$ due to our definition of $\SNR=\beta\frac{E_s}{\No}=\frac{\beta}{\No}$.
We now show that the minimizer $\sigma_\star$ of \fref{eq:hassibi_FP} coincides exactly to that fixed point solution given by state evolution.

Since $\sigma_\star$ is the unique minimal solution to $F_M(\sigma)$, $F_M'(\sigma_\star)=0$ where $F_M'(\sigma) =\frac{\dd}{\dd\sigma}F_M(\sigma)$.
Straightforward differentiation of $F_M(\sigma)$ yields
%
\begin{align*}
&\frac{\dd}{\dd\sigma}F_M(\sigma)
=
\frac{1}{2}\!\left(\frac{1}{\beta}-\frac{M-1}{M}\right)\!
- \frac{\No}{2\beta\sigma^2}
\\&\quad+
\frac{1}{M}
\sum_{k\in K}
\!\left[
\!\left(1 - \frac{k^2}{\sigma^2}
\right)\!
Q\!\left(\frac{k}{\sigma}\right)
+\frac{k}{\sqrt{2\pi\sigma^2}}
\exp\!\left( - \frac{k^2}{2\sigma^2}\right)
\right]\! \notag
.
\end{align*}
Rearranging $F_M'(\sigma_\star)=0$ results in
\begin{align}
\label{eq:sigstar}
\sigma_\star^2 =
\No + \beta
\!\left[
\frac{M-1}{M}\sigma_\star^2
+
\frac{2}{M}
\sum_{k\in K} T(\sigma_\star,k)
\right]\!,
\end{align}
where we define the shorthand notation
\begin{align*}
T(\sigma,k) =
(k^2 - \sigma^2)
Q\!\left(\frac{k}{\sigma}\right)\!
- \frac{k \sigma}{\sqrt{2\pi}}
\exp\!\left( - \frac{k^2}{2\sigma^2}\right)\!.
\end{align*}

We now show that \fref{eq:sigstar} corresponds to fixed-point solution to the SE equation $\sigma_\star^2 = \No + \beta\Psi^\textnormal{PAM}(\sigma_\star^2)$, where $\Psi^\textnormal{PAM}(\sigma^2)$ is derived in \fref{eq:PAM_PSI}.
We start by partitioning $K = K_\text{L} \cup K_M \cup K_\text{U}$ where $K_\text{L} = \{2,4,\ldots,M-2\}$, $K_\text{M} = M$, and $K_\text{U} = \{M+2,\ldots,2(M-1)\}$.
We will use the fact that $T(\sigma,0) = -\frac{1}{2}\sigma^2$. For $K_\text{L}$, we have
\begin{align*}
\sum_{k\in K_\text{L}}\! T(\sigma_\star,k) &=
\sum_{k'\in K_\text{L}}\! T(\sigma_\star,M-k')
=
\sum_{\ell = 1}^{\frac{M}{2}-1} T(\sigma_\star,M-2\ell)
\\
&=\sum_{\ell = 1}^{M/2} T(\sigma_\star,M-2\ell) +\frac{1}{2}\sigma_\star^2.
\end{align*}
For $K_\text{U}$, we have
\begin{align*}
\sum_{k\in K_\text{U}}\! T(\sigma_\star,k) &=
\sum_{k'\in K_\text{L}}\! T(\sigma_\star,M + k')
=
\sum_{\ell = 1}^{\frac{M}{2}-1} T(\sigma_\star,M+2\ell)
\\
&=\sum_{\ell = 2}^{M/2} T(\sigma_\star,M+ 2(\ell-1)),
\end{align*}
so that
\begin{align*}
\sum_{k\in K_\text{U}}T(\sigma_\star,k)  + T(\sigma_\star,M) =
\sum_{\ell = 1}^{M/2} T(\sigma_\star,M+2(\ell-1)).
\end{align*}
Therefore, the proof is complete as the RHS of \fref{eq:sigstar} is
\begin{align*}
&\frac{M-1}{M}\sigma_\star^2
+
\frac{2}{M}
\sum_{k\in K} T(\sigma_\star,k)
=
\sigma_\star^2
\\
&+\frac{2}{M}\sum_{k=1}^{M/2}
\!\left[
T(\sigma_\star,\alpha+1 - 2k)
 + T(\sigma_\star, \alpha+1 + 2(k -1))
\right]\! \notag
\\&
=\sigma_\star^2 +
\frac{2}{M}\sum_{k=1}^{M/2}
\!\left[
T(\sigma_\star,\bar\alpha_k)
 + T(\sigma_\star,\alpha_k)
\right]\!
= \Psi^\textnormal{PAM}(\sigma_\star^2).
\end{align*}
We note that the BPSK case, i.e., $M=2$, was shown in \cite{TAXH2015} and the corresponding proof for \mLAMA was given in \cite{JMS2016}.
The presented proof shows that the BOX-relaxed method in \cite{TXH2018} and \smLAMA under uniform hypercube prior achieves the same fixed-point in \fref{eq:fpequation}.
Moreover, due to the decoupling property of \LAMA detailed in \fref{sec:decoupling_section}, the symbol error-rate of real-valued $M$-PAM system is given by $2(1-M^{-1})Q(1/\sigma_\star)$.
We note that by \fref{lem:identical_separable}, our result can be generalized to that of a $M^2$-QAM systems, which was not included in the analysis provided in \cite{TXH2018}.

\section{Proof of \fref{lem:ERT_QAM}}\label{app:ERT_QAM}

By \fref{eq:qam_SE_deriv}, we can compute $\frac{\dd \Psi^\text{QAM}(\sigma^2)}{\dd \sigma^2}$ as
\begin{align*}
&\frac{\dd \Psi^\text{QAM}(\sigma^2)}{\dd \sigma^2}
=
1-\frac{1}{M} -
\frac{2}{M}
\!\left(
\frac{2\alpha
}{\sqrt{\pi\sigma^2}}e^
{
-\frac{4\alpha}{\sigma^2}
}
+ Q\!
\left(
\frac{2\alpha}{\sigma/\sqrt{2}}
\right)
\right) \notag \\
&-
\frac{2}{M}
\sum_{k=1}^{M/2-1}
\Bigg[
\frac{1}{\sqrt{\pi\sigma^2}}
\left(
\alpha_k
\exp\!\left(
{-\frac{\alpha_k^2}{\sigma^2}}
\right)
+
\bar\alpha_k
\exp\!\left(
-\frac{\bar\alpha_k^2}{\sigma^2}\right)\!
\right) \notag
\\
&+ Q\!\left(
\frac{\bar\alpha_k}{\sigma/\sqrt{2}}\right)
+ Q\!\left(
\frac{\alpha_k}{\sigma/\sqrt{2}}
\right)
\Bigg]\!.
\end{align*}

All terms on the right-hand side of $1-1/M$ are negative for $\sigma^2>0$ and attain its maximum of $0$ as $\sigma^2\to0$. Thus, we have that
\begin{align*}
\min\limits_{\sigma^2\geq0}\left\{\left(\frac{\dd\Psi^\textnormal{QAM}(\sigma^2)}{\dd\sigma^2}\right)^{\!-1}\right\} & = \lim\limits_{\sigma^2\to0}\left(\frac{\dd\Psi(\sigma^2)}{\dd\sigma^2}\right)^{\!-1} \\
& =(1-1/M)^{-1}.
\end{align*}
\revision{
In order to show that \mLAMA also recovers original signal when $\beta=\betamin$, we use \fref{lem:unique_fp_MRT} and observe that $(\dd\Psi^\textnormal{QAM}(\sigma^2_\star)/\dd\sigma^2_\star)^{-1}$ is maximized only at $\sigma^2\to0$ and hence no other $\sigma^2_\star>0$ satisfies $\betamin = (\dd\Psi^\textnormal{QAM}(\sigma^2_\star)/\dd\sigma^2_\star)^{-1}$.}

\bibliographystyle{IEEEtran}
\bibliography{bib/VIPabbrv,bib/confs-jrnls,bib/publishers,bib/cjBibTeX_210206}

\end{document}